\documentclass[%
 aip,
 amsmath,amssymb,
 reprint,%
]{revtex4-1}

\usepackage{graphicx}
\usepackage{dcolumn}
\usepackage{bm}

\usepackage[utf8]{inputenc}
\usepackage[T1]{fontenc}
\usepackage{mathptmx}

\newcommand{\Zn}{\text{Z}}

\newcommand{\Hn}{\text{H}}
\newcommand{\gn}{\text{g}}
\newcommand{\mx}{\text{max}}

\newcommand{\vv}{\text{v}}

\newcommand{\ef}{\text{eff}}

\usepackage{url}
\usepackage[version=3]{mhchem}
\usepackage{color}
\usepackage{amsmath}
\usepackage{lipsum}
\usepackage{amssymb}
\usepackage[version=3]{mhchem}
\usepackage{mathrsfs}      
\usepackage{booktabs}   
\usepackage{subfigure}
\usepackage{float}
\usepackage{appendix}
\usepackage{comment}

\usepackage{epsfig}
\usepackage{dcolumn}%
\usepackage{relsize}

\begin{document}

\title{Consistent Kinetic-Continuum Dissociation Model I: Kinetic Formulation} 

\author{Narendra Singh}
 \email{singh455@umn.edu.}
\author{Thomas Schwartzentruber}%
\affiliation{ 
Department of Aerospace Engineering and Mechanics, 
University of Minnesota, Minneapolis, MN 55455}%


\date{\today}

\begin{abstract}In this article, we propose a generalized non-equilibrium chemical kinetics model from \textit{ab initio} simulation data obtained using accurate potential energy surfaces developed recently for the purpose of studying high-temperature air chemistry.  First, we present a simple cross-section model for dissociation that captures recent \textit{ab initio} data accurately. The cross-section model is analytically integrated over Boltzmann distributions and general non-Boltzmann distributions to derive general non-equilibrium dissociation model. The general non-Boltzmann model systematically incorporates key physics such as dependence on translational energy, rotational energy, vibrational energy, internal energy, centrifugal barrier and non-Boltzmann effects such as overpopulation and depletion of high energy states. The model is shown to reproduce the rates from QCT for Boltzmann distributions of internal energy states. Reduced rates in non-equilibrium steady state due to depletion of high internal energy states are also predicted well by the model. Furthermore, the model predicts the enhanced rates as observed due to significant overpopulation of high vibrational states relative to Boltzmann distributions while the gas is in non-equilibrium in the transient phase. The model provides a computationally inexpensive way of incorporating non-equilibrium chemistry without incurring additional cost in the existing computational tools. Further comparisons of the model are carried out in another article, where simplifications to the model are proposed based on the results.
\end{abstract}

\maketitle

\section{Introduction}

 In hypersonic flows, dissociation of the gas in the regions immediately behind shock waves, is coupled to the internal energy of the gas. The gas molecules in the excited vibrational states are strongly favored for dissociation. The effect of rotational energy on dissociation is similar but is less pronounced relative to the vibrational energy. Recently, \textit{ab initio} methods such as direct molecular simulations (DMS) \cite{valentini2016dynamics,grover2019direct}, master-equation analysis \cite{panesi2014pre,macdonald2018_QCT, andrienko2018vibrational,magin2012coarse} and quasi-classical trajectory calculations (QCT) \cite{chaudhry2018qct,voelkel2017multitemperature}  have quantified the importance of ro-vibrational energy in the state-specific dissociation rates of air species. This has been possible due to recently developed accurate potential energy surfaces (PESs) for N$_2$-N$_2$ and N–N$_2$ collisions  \cite{paukku2013global,paukku2014erratum}, O$_2$–O$_2$ \cite{paukku2017potential} and O-O$_2$ \cite{O2OTruhlar} collisions, N$_2$-O$_2$ collisions \cite{varga2016potential}, and N$_2$–O collisions \cite{lin2016global} for the purpose of studying the air chemistry relevant to hypersonic flows.
 
 Dissociating gas in the shock heated regions is in non-equilibrium. 
 Due to strong favoring of high internal energy states, dissociation removes the molecules in the high internal energy states. These high energy states are then repopulated by vibration-translation and rotation-translation energy exchanges via non-reactive collisions. The balance of dissociation and re-population results in a quasi-steady state (QSS). In the QSS phase, the gas has depleted internal energy distributions relative to corresponding equilibrium (Boltzmann) distribution \cite{valentini2016dynamics,macdonald2018_QCT,andrienko2018vibrational}. The depletion of molecules in the high ro-vibrational states results in reduced dissociation rates (by a factor of $3-5$ for nitrogen \cite{valentini2016dynamics,macdonald2018construction_DMS} and $2-5$ for oxygen\cite{grover2019jtht}) relative to Boltzmann distribution based estimates \cite{bender2015improved,chaudhry2019statistical}. Contrary to the depletion effects, overpopulation of high internal energy states occurs in the early phase of excitation of ro-vibrational energy. High energy non-reactive collisions, result in overpopulation of high internal energy states compared to the corresponding Boltzmann distribution \cite{sahai2019flow}. This overpopulation  enhances the rates of dissociation, relative to Boltzmann distributions.

For accurate prediction of dissociation, the dissociation models in computational fluid dynamics (CFD) solvers should include the aforementioned coupling of dissociation to internal energy. To incorporate internal energy effects in the development of continuum dissociation rates,  Arrhenius-like descriptions have been extended by introducing separate temperatures for rotational energy ($T_{rot}$) and vibrational energy ($T_v$). For instance, the widely used Park model \cite{park1988two,park1993review, park1989assessment} uses an effective temperature ($\sqrt{T T_v}$) in Arrhenius rates, where parameters are empirically adjusted to fit experimental results \cite{byron1966shock,appleton1968shock,hanson1972shock}, which have uncertainties (about an order of magnitude for nitrogen dissociation rates). Using existing state-specific kinetic dissociation rates, \textit{continuum} dissociation models \textcolor{red}{\cite{marrone1963chemical,knab1995theory}} have been developed, but these are not consistent with recent  \textit{ab initio} state-specific data. Existing \textit{kinetic} rate models (or their modifications) have been refitted to the recent \textit{ab initio} data \cite{kustova2016advanced,gimelshein2017modeling,wadsworth1997vibrational} and more elaborate fits to new state-specific dissociation rates, consistent with \textit{ab initio} data have also been proposed \cite{luo2017ab}. However, a model that is analytically consistent between kinetic and continuum descriptions (suitable for large-scale CFD), that is consistent with \textit{ab initio} data, has not yet been developed. 


In parallel efforts, master equation approach has been used to study systems relevant to air chemistry, for instance, N$_2$+O$_2$ \cite{andrienko2018vibrational}, and N$_2$+N$_2$\cite{panesi2013rovibrational}. 
Master equation requires state-specific transition rates and reaction rates (corresponding to the pre-collision distribution) which are computed using quasi-classical trajectory (QCT) calculations   \cite{truhlar1979atom}.
Due to the high dimensionality of state space ($\approx 10^{15}$ transition rates for N$_2$+N$_2$), energy state binning (averaging) assumptions are required \cite{macdonald2018_QCT,macdonald2018construction_DMS}, and maintaining desired accuracy is still an open question\cite{macdonald2018_QCT}. 


 In the current work, we present both continuum (rate-based) and kinetic (state specific rate-based) rate models that are analytically consistent and are formulated based on quantum chemistry predictions. We first identify and express the dependence of the kinetic rates on key dominant physics such as vibrational energy, rotational energy and relative translational energy based on \textit{ab-initio} data. We then present continuum based dissociation rate constants which incorporate non-Boltzmann effects using suprisal analysis based model by Singh and Schwartzentruber \cite{Singhpnas, Singhpnasplus}, also based on \textit{ab-initio} data. The average energy of dissociating molecules is also derived, which is necessary to accurately track the evolution of average vibrational energy.  The model accurately captures the key physics observed in ab-intio calculations and, since it is analytical, the relative influence of each set of physics is contained in different terms and can be
analyzed. For instance, we report the dissociation rate enhancement factors due to overpopulation of high internal energy states, and the reduction in the dissociation rates due to depletion of high internal energy states.  


 \section{Kinetic to Continuum Framework}{\label{master_eqn_framework}}
In this section, we outline the main equations that link a kinetic description of internal energy relaxation and dissociation to the most widely-used continuum framework. 

Consider a collision where diatomic molecule $AB (j,v)$ interacts with a partner $C$. leading to internal energy exchange or dissociation,
 \begin{equation}
 \begin{split}
     AB (j,v) + C \xrightarrow{k_{jv-j'v'}}  AB(j',v')+C \\
     AB (j,v) + C \xrightarrow{k_{jv-d}}  A+B+C
     \end{split}
 \end{equation}
Here $(j,v$) represents the rovibrational state of the molecule and $C$ can be either a diatom or atom. 

In principle, such kinetic rate expressions can be used to solve the master equation where, if all transition rates between all rovibrational states are input, the population of each rovibrational state can be computed. However, this quickly becomes computationally intractable, for example there are approximately $10^{15}$ such transitions for N$_2$-N$_2$ collisions and $10^{7}$ for N-N$_2$ collisions. In order to derive a tractable model for the evolution of rovibrational populations in a gas, one must group (or ``coarse-grain'') the full set of quantized rovibrational states \cite{panesi2014pre}. In this manner, only the transition rates between rovibrational energy \emph{groups} are required. 
For example, by combining rotational energy states into a single group, characterized by the average rotational energy ($\langle \epsilon_{rot} \rangle$), an evolution equation for the vibrational energy populations, $AB(v)$ at a given translational temperature ($T$), can be written as,
   \begin{equation}
   \begin{split}
    \frac{ d\ [AB(v)] }{dt} =  \sum_{v'\neq v} k_{v'-v}  [AB(v')][C] 
    - \left ( \sum_{v'\neq v} k_{v-v'} \right)[C] [AB(v)] \\ -  k_{v-d} [C] [AB(v)] 
      +[\text{recomb}.]
     \label{master_eqn}
     \end{split}
 \end{equation}
 
Here, $k_{i-i'} $ is the rate constant for transitioning from state $i$ to $i'$ during a collision with partner specie $C$, $d$ denotes the dissociated state, and $[\text{recomb}.]$ is a term containing the production of specific $AB(v)$ states due to recombination reactions. $k_{i-i'} (\equiv k_{i-i'} (T,\langle \epsilon_{rot} \rangle))$ depends on $T$ and $\langle \epsilon_{rot} \rangle$ but for notational brevity, we have dropped the dependence. Note that, in Eq.~\ref{master_eqn}, the rate constants are  averaged over the rovibrational states of the colliding partner when it is a diatom. This assumption is justified later in section 3. Furthermore, since the current work focuses on dissociating flows where recombination is negligible, we ignore recombination while developing our model (recombination is addressed later in section  VI of subsequent article Ref.~\cite{SinghCFDII}).

In order to obtain a continuum expression, the moment of Eq.~\ref{master_eqn} is taken in order to derive an equation for the evolution of average vibrational energy, $\langle \epsilon_v \rangle$,
  \begin{equation}
  \begin{split}
    \frac{ d \langle \epsilon_v \rangle}{dt} = \sum_v  \sum_{v'\neq v} k_{v'-v} f(v')[C]  \epsilon_v(v)
    -\sum_v  \left ( \sum_{v'\neq v} k_{v-v'} \right)[C] f(v) \\- k_{AB-C} [C] \left( \langle \epsilon_v^d \rangle - \langle \epsilon_v \rangle) \right)
     \label{master_eqn_vib2}
  \end{split}
 \end{equation}
Here, $f(v)$ is the distribution function of vibrational energy states, and $\epsilon_v(v)$ is the vibrational energy associated with a specific $v$ state. Furthermore, in the final term, $k_{AB-C}$ is the overall dissociation rate and $\langle \epsilon_v^d \rangle$ 
is the average vibrational energy of the dissociating molecules. The equations for $k_{AB-C}$ and $\langle \epsilon_v^d \rangle$ will be presented and discussed shortly.

First, the Landau-Teller (LT) model is used to simplify the first two terms. The LT equation can only be derived from the master equation under certain assumptions \cite{nikitin200870,vincenti1965introduction}, such as mono-quantum transitions. However, use of the LT model is common in CFD codes used for flows in thermal nonequilibrium, and there is evidence that vibrational energy relaxation time constants, determined both experimentally and computationally, can be accurately fit with the LT model \cite{valentini2016dynamics,grover2019jtht,grover2019direct}. Since the current work focuses on developing a dissociation model, the LT expression is used to determine the evolution of the average vibrational energy. However, it is important to note, that the current work does not assume equilibrium (Boltzmann) distributions based on this average, rather a recently developed model for non-Boltzmann vibrational energy distributions \cite{Singhpnas, Singhpnasplus} is used (described later in section ~\ref{Non_Boltz_Distro_Int}). The equations for average vibrational energy and concentration of species, $[AB]$, become,
 
 \begin{equation}
 \begin{split}
    \frac{ d \langle \epsilon_v \rangle}{dt} = \frac{\langle \epsilon_v^* \rangle-\langle  \epsilon_v \rangle}{\tau_{\text{mix}}} -k_{AB-AB} [AB] (\langle \epsilon_v^d \rangle - \langle \epsilon_v \rangle) \\
    - k_{AB-C} [C] (\langle \epsilon_v^d \rangle - \langle \epsilon_v \rangle) ~,
     \label{LandauTeller_modified_general}
\end{split}
 \end{equation}
 
  \begin{equation}
    \frac{ d [AB]}{dt} = -k_{AB-C} [AB][C]~,
     \label{Rate_eqn_general}
 \end{equation}
where $\tau_{\text{mix}}$ is the relaxation time constant of the mixture (representing both diatom-diatom and atom-diatom collisions).
This article focuses on nitrogen gas, for which Eqs.~\ref{LandauTeller_modified_general} and \ref{Rate_eqn_general} are,
  \begin{equation}
  \begin{split}
    \frac{ d \langle \epsilon_v \rangle}{dt} = \frac{\langle \epsilon_v^* \rangle-\langle  \epsilon_v \rangle}{\tau_{\text{mix}}} -k_{N_2-N_2} [N_2] (\langle \epsilon_v^d \rangle - \langle \epsilon_v \rangle) \\ - k_{N_2-N} [N] (\langle \epsilon_v^d \rangle - \langle \epsilon_v \rangle) ~,
     \label{LandauTeller_modified}
     \end{split}
 \end{equation}
  \begin{equation}
    \frac{ d [N_2]}{dt} = -k_{N_2-N_2} [N_2][N_2] - k_{N_2-N} [N_2][N] ~.
     \label{Rate_eqn_pop}
 \end{equation}
Equations \ref{LandauTeller_modified} and \ref{Rate_eqn_pop} are source-terms in the vibrational energy equation and species conservation equation that are commonly used in hypersonic CFD codes. This article proposes models for the two important quantities; the dissociation rate constant ($k$) and the average energy of dissociating molecules ($\langle \epsilon_v^d \rangle$), which are both dependent on the local nonequilibrium \textit{state}\footnote{the word `state' is used in loose manner unlike thermodynamic state where it is well defined)} characterized by $T$, $\langle \epsilon_v \rangle$ and $\langle \epsilon_{rot} \rangle$.  

Upon taking the moment of Eq.~\ref{master_eqn} to obtain Eq.~\ref{master_eqn_vib2}, the overall dissociation rate coefficient is derived as \cite{truhlar1979atom,bender2015improved},
\begin{widetext}
\begin{equation}
\begin{split}
   k(T,  \langle \epsilon_{rot} \rangle, \langle \epsilon_v \rangle ) = \frac{1}{S} \left(\frac{ 8 k_B T }{\pi \mu_C  }\right)^{1/2} \pi b_{\max}^2  \sum_v^{v_{\mx}} \sum_j^{j_{\mx}(v)} \hspace{2in}
  \\
   \times  \int _0^\infty  p (d\ |\ \epsilon_{rel},\epsilon_{v},\epsilon_{rot}) \left(\frac{\epsilon_{rel}}{k_B T}\right) \exp\left[- \frac{\epsilon_{rel}}{k_B T} \right] d \left(\frac{\epsilon_{rel}}{k_B T}\right) f^{NB}(j,v) \hspace{1.2in},
  \end{split}
  \label{Eratemain}
\end{equation}

and the average energy of dissociated molecules ($\langle \epsilon_{v}^d \rangle$), is derived as, 

\begin{equation}
    \langle \epsilon_{v}^d \rangle = \cfrac{\sum_v^{v_{\mx}} \sum_j^{j_{\mx}(v)} \int _0^\infty  \epsilon_{v}p (d\ |\ \epsilon_{rel},\epsilon_{v},\epsilon_{rot}) \left(\frac{\epsilon_{rel}}{k_B T}\right) \exp\left[- \frac{\epsilon_{rel}}{k_B T} \right] d \left(\frac{\epsilon_{rel}}{k_B T}\right) f^{NB}(j,v)}{\sum_v^{v_{\mx}} \sum_j^{j_{\mx}(v)} \int _0^\infty p (d\ |\ \epsilon_{rel},\epsilon_{v},\epsilon_{rot}) \left(\frac{\epsilon_{rel}}{k_B T}\right) \exp\left[- \frac{\epsilon_{rel}}{k_B T} \right] d \left(\frac{\epsilon_{rel}}{k_B T}\right)f^{NB}(j,v)}~.
    \label{Edv_eqn}
\end{equation}
\end{widetext}
In Eqs. \ref{Eratemain} and \ref{Edv_eqn},  $\epsilon_{rel}$ is the relative translational energy of a collision pair (the kinetic expressions have been integrated over a Maxwell-Boltzmann velocity distribution function), $k_B$ is the Boltzmann constant, $\mu_C$ is the reduced mass of the collision pair, and $S$ is a the symmetry factor ($S=2$ for diatom-diatom collisions and $S=1$ for atom-diatom collisions). 

In Eqs. \ref{Eratemain} and \ref{Edv_eqn}, the two expressions that must be modeled are (i) the probability of dissociation given the molecular state, $p (d\ |\ \epsilon_{rel},\epsilon_{v},\epsilon_{rot})$, and (ii) the non-Boltzmann distribution of internal energy states, $f^{NB}(j,v)$. As discussed in the next section, the probability expression corresponds to an average over a range of impact parameters, where $b_{max}$ is the maximum impact parameter, and where $\epsilon_{v}$ and $\epsilon_{rot}$ are functions of the $(j,v)$ state.

\section{Reaction Cross-Section Model} \label{RCS}
In this section, a simple analytical expression for the probability of dissociation is proposed (equivalent to the dissociation cross-sections normalized by a reference cross-section), and justification for the functional form of each term is described in detail. The proposed dissociation probability model is:
\begin{widetext}
\begin{equation}
\begin{split}
  p (d\ |\ \epsilon_{rel},\epsilon_{v},\epsilon_{rot}) = C_1 \overbrace{\left[\frac{\epsilon_{rel}+\epsilon_{int}-\epsilon_d^{\ef} }{\epsilon_{d}}\right]^{\alpha} \frac{\epsilon_{d}}{\epsilon_{rel}}}^{\text{I : Collision Energy}} 
  \overbrace{\exp\left[\beta \frac{\epsilon_{rot}^{\text{eff}}}{\epsilon_d}\right]}^{\text{II: Rotational Energy}} \hspace{1.5in}
  \\
  \times \underbrace{ \exp\left[\gamma \frac{\epsilon_{v}}{\epsilon_d}\right]}_{\text{III: Vibrational Energy}} \underbrace{\exp\left[\delta \frac{|\epsilon_{int}-\epsilon_d|}{\epsilon_d}\right]}_{\text{IV: Internal Energy}} \hspace{0.25in} \epsilon_{rel}+\epsilon_{int} \geq\epsilon_d^{\ef}
  \\
  = 0 \hspace{3.35in} \text{otherwise }
  \end{split}
    \label{prob_expression_main}
\end{equation}
\end{widetext}
In this expression, $\epsilon_{rot}^{\text{eff}}= \epsilon_{rot}-(\epsilon_d^{\text{eff}}-\epsilon_d) =\epsilon_{rot}- \theta_{CB}(\epsilon_{rot}) $ is the effective rotational energy that includes the centrifugal energy barrier effect ($\theta_{CB}$), and similarly, $\epsilon_d^{\text{eff}}=\epsilon_d+\theta_{CB}(\epsilon_{rot})$) is the effective dissociation energy. The parameters $C_1, \alpha, \beta, \gamma$ and $\delta$, are model parameters that are determined by studying cross-section data from recent ab-initio calculations. The probability model in Eq.~\ref{prob_expression_main} can be directly used in DSMC, for example instead of the total collision energy model (TCE), however, DSMC modeling is not the focus of the current article. 

The purpose of the proposed model is to capture the most dominant reaction cross-section trends with a \textit{simple} functional form that can be \textit{analytically} integrated. Therefore, all of the terms in Eq. \ref{prob_expression_main} are neither entirely unique, nor the most exact. Indeed, the cross-sections obtained through ab-initio calculations could be fit using polynomial or Fourier-series basis functions, or potentially by neural networks \cite{koner2019exhaustive}. However, we find very clear trends in the ab-initio data, for which elaborate functional fits are not necessary; rather, simplified functions are accurate for the purpose of kinetic and continuum engineering analysis.

The ab-intio data used in this article comes from the quasi-classical-trajectory (QCT) simulations performed by Bender \textit{et al.} \cite{bender2015improved}. The QCT data are obtained by sampling relative translational and rotational energies from five different temperatures ranging from $T=8,000$ K to $T=30,000$ K, assuming Maxwell-Boltzmann distributions. For each $T$, vibrational energies are sampled from a Boltzmann distribution at $T_v$, which also ranges from $8, 000$ K to $30, 0000$ K. Overall 2.4 billion trajectories have been calculated. For QCT simulation details, including trajectory integration, impact parameters, Maxwell-Boltzmann distribution sampling, uncertainty estimates, and post-processing, readers are referred to the article by Bender \textit{et al.} \cite{bender2015improved}. Construction of the model in Eq. \ref{prob_expression_main} was guided by the trends observed in the QCT data, which are described in the following sub-sections.

\subsection{Dependence on collision partner internal energy}

\begin{figure}
  \centering
    \includegraphics[width=3.2in]{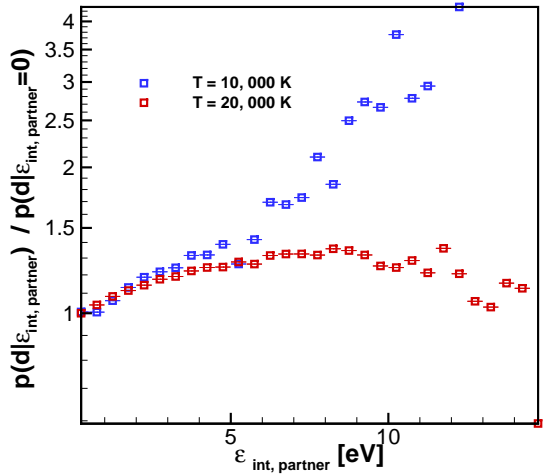}
   \label{prob_QCT_v}
   \caption{Dissociation probability as function of internal energy of the colliding partner ($p(d|\epsilon_{int,partner}$))}
   \label{prob_QCT_eint}
 \end{figure}

The dependence of the dissociation probability of a molecule on the collision partner's internal energy is shown in Fig.~\ref{prob_QCT_eint}.
Specifically, the relative translational and internal energies of the molecules undergoing dissociation are sampled from Boltzmann distributions (at $T = 10, 000$ K and $T= 20, 000$ K). The variation in the dissociation probability is plotted for a wide range of collision partner internal energies, based on a large number of QCT calculations. More specifically, the reaction probability is integrated over all parameters except the internal energy of the colliding partner, to obtain an expression for $p(d|\epsilon_{\text{int,partner}})$,
\begin{widetext}
\begin{equation}
\begin{split}
   p(d|\epsilon_{\text{int,partner}}) \hspace{4 in} \\ =  \sum_v^{v_{\mx}} \sum_j^{j_{\mx}(v)}  \int _0^\infty  p (d\ |\ \epsilon_{rel},\epsilon_{v},\epsilon_{rot},\epsilon_{\text{int,partner}}) \left(\frac{\epsilon_{rel}}{k_B T}\right) \exp\left[- \frac{\epsilon_{rel}}{k_B T} \right] d \left(\frac{\epsilon_{rel}}{k_B T}\right) f(j,v; T)
  \\
  \end{split}
  \label{coll_partner_energy}
\end{equation}
\end{widetext}

As seen in Fig.~\ref{prob_QCT_eint}, the variation in probability is at most 4x. In contrast, as shown in upcoming sections, the dependence on the dissociating molecule's internal energy spans many orders of magnitude. Figure~\ref{prob_QCT_eint} indicates that although a high value of the partner's internal energy increases the effective reaction cross-section, by providing additional energy towards crossing the dissociation barrier, the internal energy of the colliding partner has a negligible effect on the overall rate constant; a conclusion also made in other recent studies \cite{chaudhry2018qct}. Therefore, a specific dependence on the collision partner internal energy is not proposed in our model (Eq.~\ref{prob_expression_main}).  

\subsection{Dependence on translation energy}
\begin{figure}
        \subfigure[]
  {
    \includegraphics[width=3.2in]{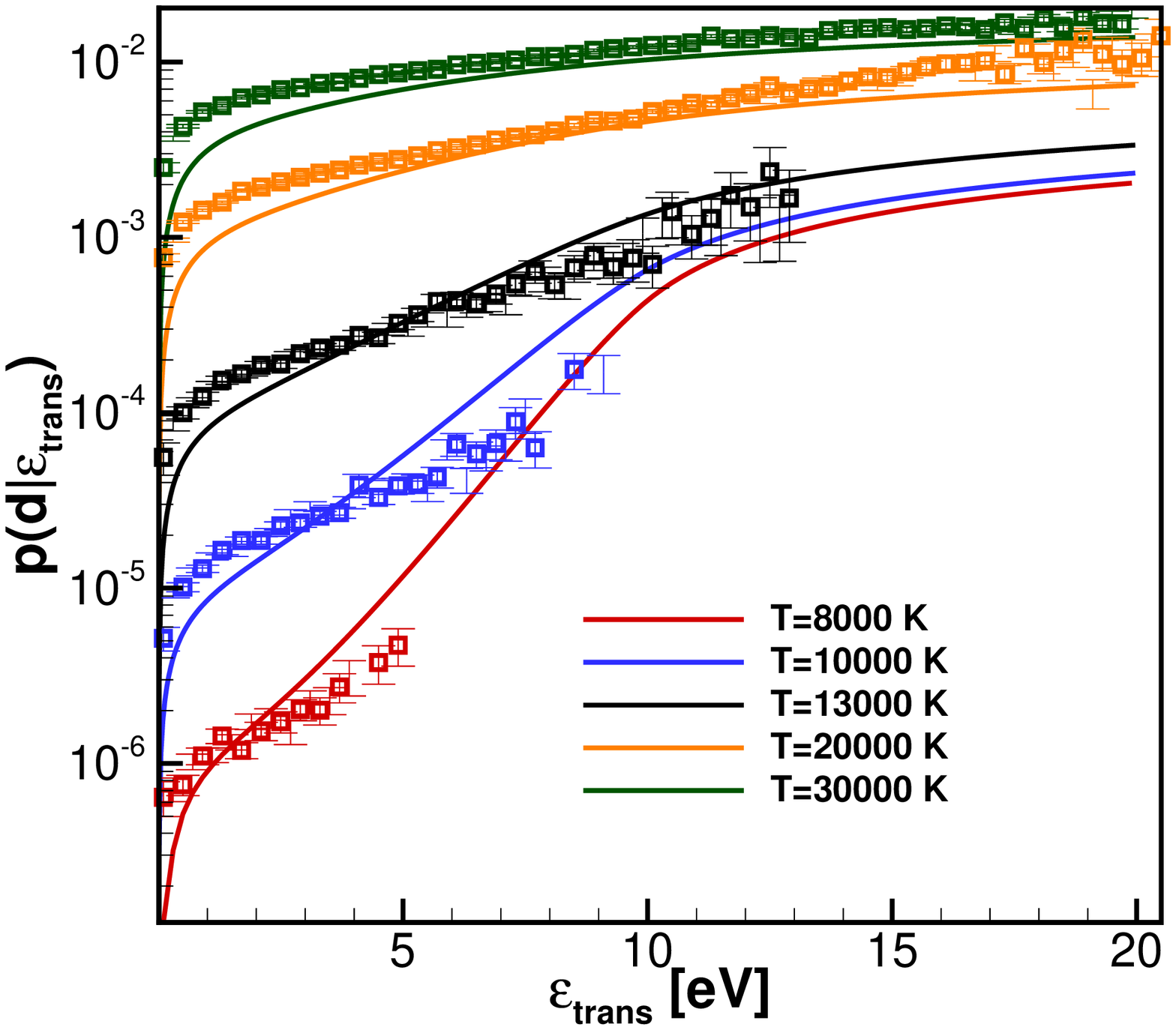}
     \label{etrans_prob}
   }
    \subfigure[]
  {
    \includegraphics[width=3.2in]{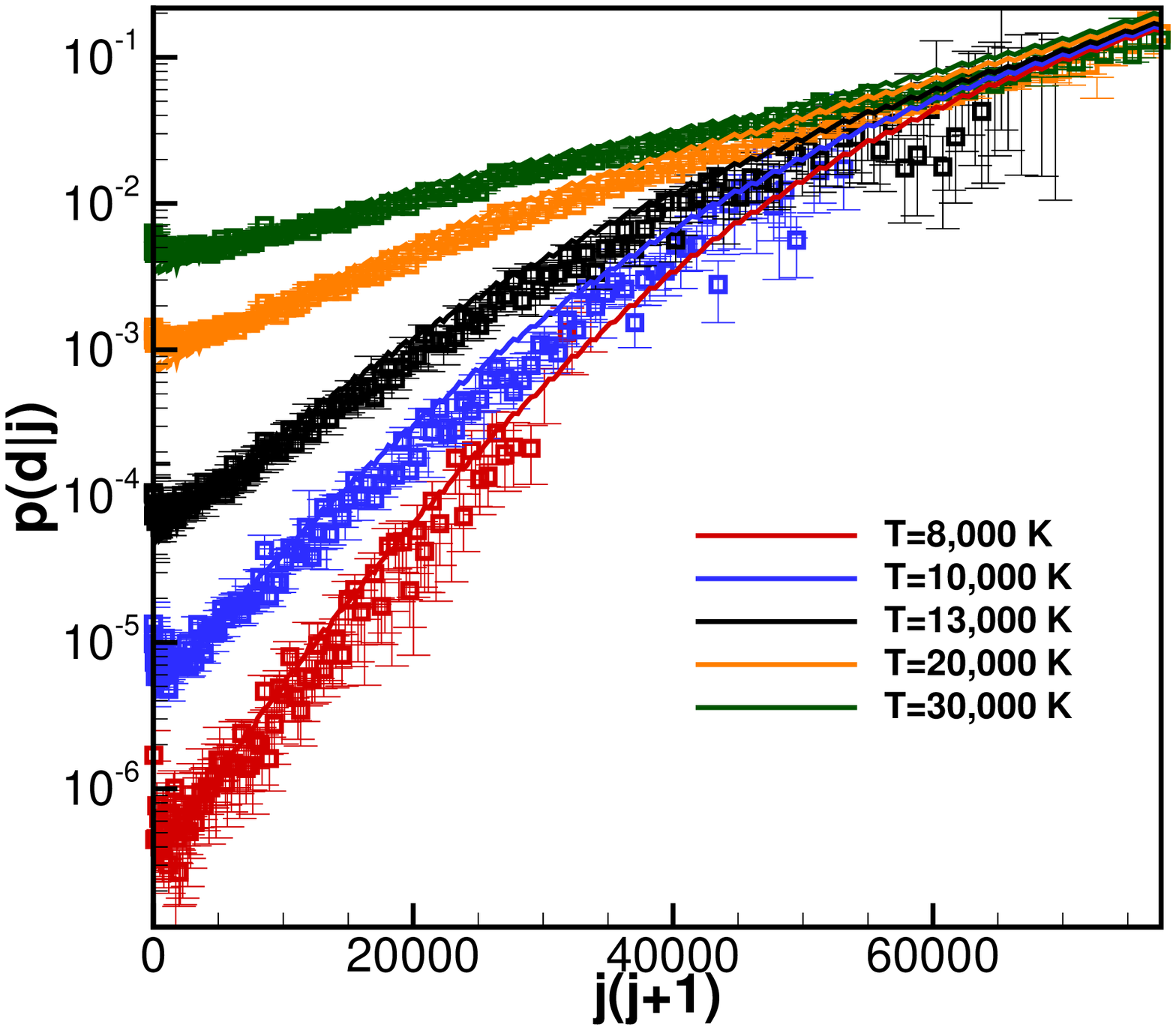}
     \label{erot_prob}
   }  
     \subfigure[]
  {
    \includegraphics[width=3.2in]{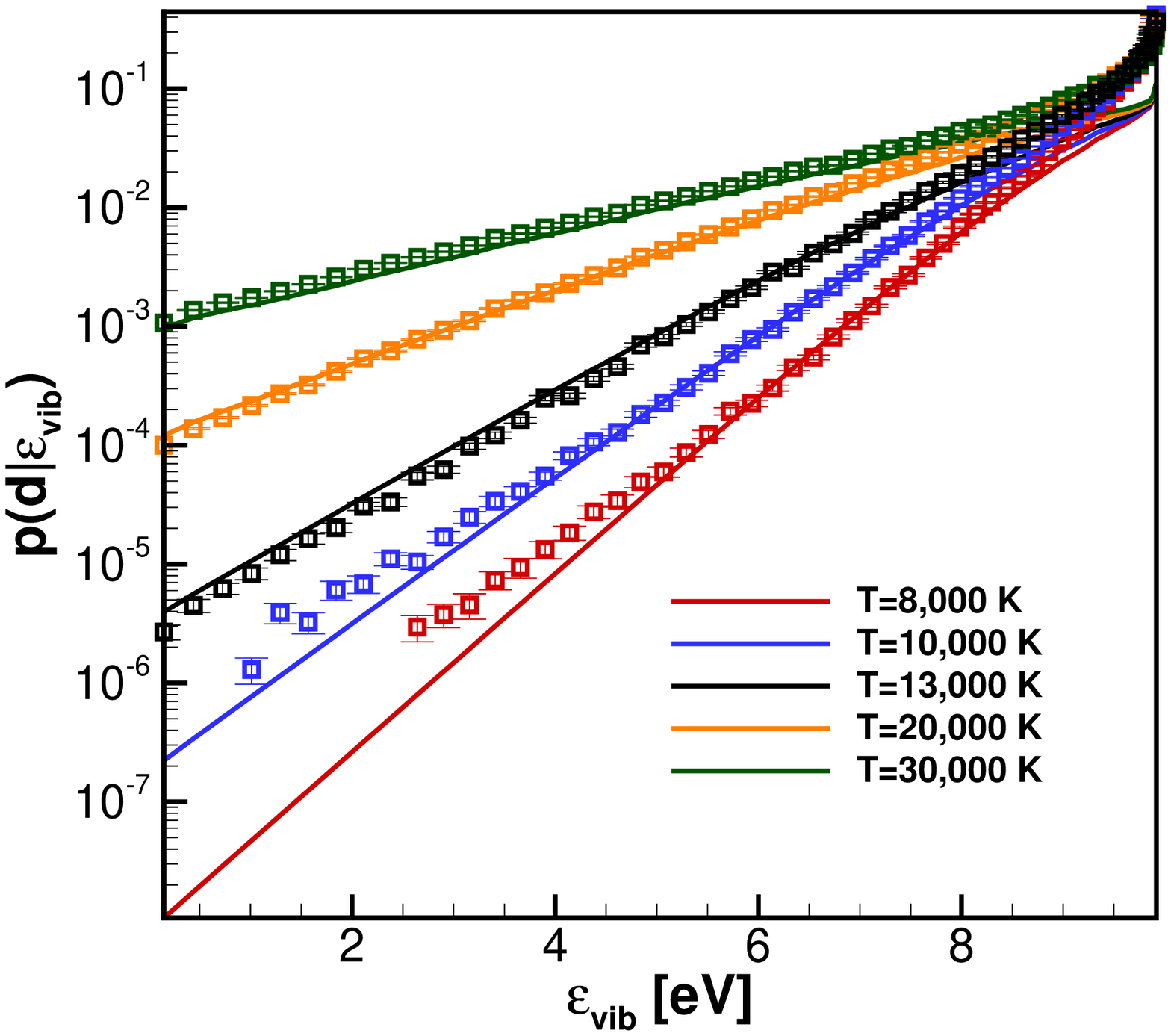}
   \label{evib_prob}
   } 
   \caption{Conditional reaction probabilities as a function of the (a) translational, (b) rotational, and (c) vibrational energy. Symbols denote QCT results \cite{bender2015improved}, and solid lines represent model results.}
   \label{prob_QCT_trans}
 \end{figure}

The first term (I) in Eq.~\ref{prob_expression_main} is proportional to the excess energy in a collision relative to the dissociation energy. The term $(\epsilon_d/\epsilon_{rel})$ ensures that in the limit of high translational energy, fly-by collisions act to decrease the reaction cross-section. Numerous collision induced dissociation models, developed using statistical considerations, have a form similar to this first term (examples include Refs. \cite{levine1971collision,maier1964dissociative,parks1973collision}). Figure~\ref{etrans_prob} plots the conditional probability of dissociation ($p(d|\epsilon_{trans})$) from a given relative translational energy, where the internal energies are sampled from Boltzmann distributions at $T$. Formally, the equation for $p(d|\epsilon_{trans})$ is
\begin{equation}
\begin{split}
   p(d|\epsilon_{trans}) =  \sum_v^{v_{\mx}} \sum_j^{j_{\mx}(v)} p (d\ |\ \epsilon_{rel},\epsilon_{v},\epsilon_{rot}) f(j,v; T)~.
  \\
  \end{split}
  \label{pdtranseqn}
\end{equation}
As shown in Fig.\ref{etrans_prob}, for high values of $T$ (corresponding to high internal energy), the influence of translational energy on the dissociation probability becomes progressively weaker. Using a single constant value of $\alpha$ (see Table \ref{parameters} in Sec. V (A)), the model (Eq. \ref{prob_expression_main}) is able to accurately reproduce the QCT data.

\subsection{Dependence on rotational energy}
Figure~\ref{erot_prob} plots the conditional probability of dissociation ($p(d|j)$) from a given rotational energy, where the relative translational energies and vibrational energies are sampled from Maxwell-Boltzmann distributions at $T$. Specifically, the equation for $p(d|j)$ is:
\begin{widetext}
\begin{equation}
\begin{split}
  p(d|j) =  \sum_{v=0}^{v_{\mx}}  \int _0^\infty  p (d\ |\ \epsilon_{rel},\epsilon_{v},\epsilon_{rot}) \left(\frac{\epsilon_{rel}}{k_B T}\right) \exp\left[- \frac{\epsilon_{rel}}{k_B T} \right] d \left(\frac{\epsilon_{rel}}{k_B T}\right) f(j,v; T)
  \\
  \end{split}
  \label{pdroteqn}
\end{equation}
\end{widetext}
As evident in Fig. \ref{erot_prob}, the dissociation probability is strongly dependent on the rotational energy of the dissociating molecule, as there is an increase of approximately four orders of magnitude as the rotational energy is varied for $T = 10, 000$ K. At lower gas temperatures, the dependence is stronger as rotational energy becomes necessary to cross the required threshold for dissociation \cite{bender2015improved}. Furthermore, Macdonald \textit{et. al.} \cite{macdonald2018_QCT}, using grouped master equation analysis, have shown that rotational energy accounts for nearly $40 \%$ of the energy required for dissociation. Therefore, to incorporate strong rotational coupling to dissociation, we propose an exponential dependence on the rotational energy ($\propto j(j+1)$) for the second term (II) in Eq. \ref{prob_expression_main}. 

In most models, the centrifugal barrier is incorporated by only modifying the dissociation energy threshold to be the effective dissociation energy ($\epsilon_d^{\text{eff}}$) \cite{pritchard1976atoms}. We propose to incorporate the centrifugal barrier related to rotational energy term along with the collision energy term (I). Although, during a collision, the rotational energy may increase, decrease, or may be ill-defined (due to its coupling with vibrational energy), the proposed form is found to be more accurate than merely updating the threshold dissociation energy.   
As shown in Fig.~\ref{erot_prob}, by using a single constant value of $\beta$ (see Table \ref{parameters} in Sec. V (A)) that controls the exponential dependence on rotational energy, the model (Eq. \ref{prob_expression_main}) is able to accurately reproduce the QCT data.
Therefore, for a given pre-collision rotational energy, a collision can result in dissociation even if the collision energy is less than $\epsilon_d^{\text{eff}}$, as long as it is greater than than $\epsilon_d$.

\subsection{Dependence on vibrational energy}
Figure~\ref{evib_prob} plots the conditional probability of dissociation ($p(d|v)$) from a given vibrational energy, where the relative translational energies and rotational energies are sampled from Maxwell-Boltzmann distributions at $T$. Specifically, the equation for $p(d|v)$ is:
\begin{widetext}
\begin{equation}
\begin{split}
  p(d|v) =  \sum_{j=0}^{j_{\mx}}  \int _0^\infty  p (d\ |\ \epsilon_{rel},\epsilon_{v},\epsilon_{rot}) \left(\frac{\epsilon_{rel}}{k_B T}\right) \exp\left[- \frac{\epsilon_{rel}}{k_B T} \right] d \left(\frac{\epsilon_{rel}}{k_B T}\right) f(j,v; T)
  \\
  \end{split}
  \label{pdvibeqn}
\end{equation}
\end{widetext}
As evident in Fig.~\ref{evib_prob}, the dissociation probability is even more dependent on the vibrational energy of the dissociating molecule compared to rotational energy. Specifically, $p(d|v)$ is  exponential and varies nearly six orders of magnitude for conditions of $T = 10, 000$ K. 
Therefore, the probability of dissociation depends most strongly on vibrational energy as also observed in recent QCT studies \cite{bender2015improved,macdonald2018_QCT}. As shown in Fig.~\ref{evib_prob}, by using a single constant value of $\gamma$ (see Table \ref{parameters} in Sec. V (A)) that controls the exponential dependence on vibrational energy, the third term (III) in the model (Eq. \ref{prob_expression_main}) is able to accurately reproduce the QCT data. As described in upcoming sections, the relative importance of vibration versus rotation is reflected in the model parameters (i.e. $\gamma > \beta$). 
A number of previous studies have also incorporated exponential dependence on vibrational energy in dissociation models \cite{kiefer1975preference,rebick1973collision,kafri1976comment}, in the context of vibrationally state-resolved rate constants. For example, Blais and Truhlar \cite{blais1977monte} proposed $ p (d\ |\ \epsilon_{rel},\epsilon_{v},\epsilon_{rot}) = B (\epsilon_{rot},\epsilon_{c}) \exp(\lambda \epsilon_v/\epsilon_d)$ for Ar+H$_2$ based on QCT calculations in \cite{blais1976monte}.

\subsection{Dependence on internal energy and quasi-bound effects}
  \begin{figure}
  \subfigure[]
  {
    \includegraphics[width=3.2in]{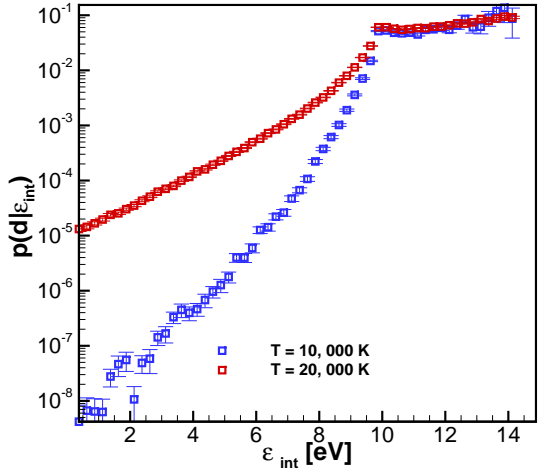}
   \label{prob_eint_QCT}
   }  
    \subfigure[]
     {
    \includegraphics[width=3.2in]{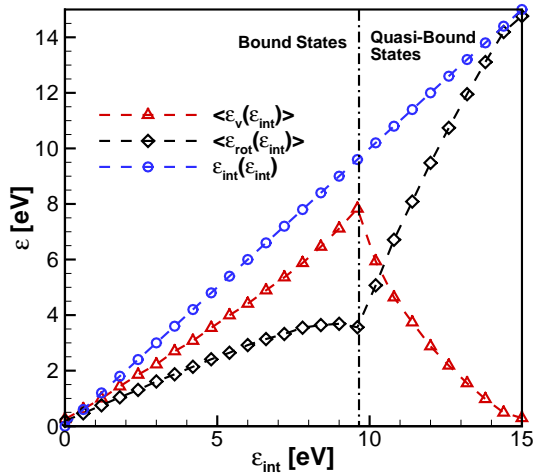}
     \label{eint_erot_evib}
   }
       \subfigure[]
  {
    \includegraphics[width=3.2in]{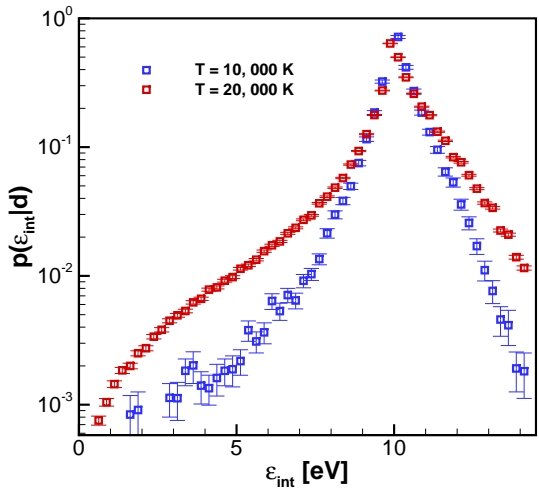}
     \label{prob_pop_QCT}
   }  
   \caption{(a) Conditional reaction probabilities as a function of internal energy, (b)  For N$_2$ molecule, vibrational ($\langle{\epsilon_{v}}(\epsilon_{int})\rangle$) and rotational ($\langle{\epsilon_{rot}}(\epsilon_{int})\rangle$)  energies as described in Eqs.~\ref{ev_int} and \ref{erot_int} respectively, (c) Reaction probability as a function of internal energy conditioned on dissociation.}
   \label{prob_QCT_trans}
 \end{figure}

It has been calculated that nearly $60\%$ of dissociation in nitrogen occurs from quasi-bound molecules for temperatures ranging from $8, 000$ K to $30, 000$ K \cite{bender2015improved}. The dissociation in bound molecules relies on translational energy to cross the dissociation barrier, while  mere redistribution or de-excitiation of internal energy is sufficient to cause dissociation in quasi-bound molecules. 
In Fig~\ref{prob_eint_QCT}, the conditional reaction probability is plotted against the internal energy of the molecule undergoing dissociation. The dependence on the internal energy (equivalently on $j,v$) can be obtained from following equation:
\begin{equation}
\begin{split}
  p(d|\epsilon_{int}) =   \int _0^\infty  p (d\ |\ \epsilon_{rel},\epsilon_{v},\epsilon_{rot}) \left(\frac{\epsilon_{rel}}{k_B T}\right) \exp\left[- \frac{\epsilon_{rel}}{k_B T} \right] d \left(\frac{\epsilon_{rel}}{k_B T}\right) 
  \\
  \end{split}
  \label{pdvibeqn}
\end{equation}
 
 As expected, $p(d|\epsilon_{int})$ varies exponentially with internal energy. The magnitude of the probability is higher for quasi-bound molecules $(\epsilon_{int} > \epsilon_d)$ compared to bound molecules $(\epsilon_{int} \leq \epsilon_d)$. However, the dependence of the probability on internal energy is stronger (steeper slope) for bound molecules than for quasi-bound molecules. This trend can be understood by considering the average increase in vibrational energy, $\langle \epsilon_{v}(\epsilon_{int})\rangle$, and rotational energy, $\langle \epsilon_{rot} (\epsilon_{int})\rangle$, corresponding to changes in internal energy for both bound and quasi-bound molecules; $(\epsilon_{int}-\Delta \epsilon_{int}/2, \epsilon_{int}+\Delta \epsilon_{int} /2)$. 
 Formally, $ \langle \epsilon_{v}(\epsilon_{int}) \rangle$ and $ \langle \epsilon_{rot} (\epsilon_{int}) \rangle$ are defined as follows:
\begin{equation}
   \langle{\epsilon_{v}}(\epsilon_{int})\rangle = \frac{\sum _{j',v'} \epsilon_{int} (0,v') I_{\Delta \epsilon_{int}} (j',v')}{\sum _{j',v'} I_{\Delta \epsilon_{int}} (j',v')};
   \label{ev_int}
\end{equation}
\begin{equation}
    \langle\epsilon_{rot}(\epsilon_{int})\rangle  = \frac{\sum _{j',v'} (2j'+1)(\epsilon_{int} (j',v')-\epsilon_{int} (0,v')) I_{\Delta \epsilon_{int}} (j',v')}{\sum _{j',v'} (2j'+1) I_{\Delta \epsilon_{int}} (j',v')}~,
    \label{erot_int}
\end{equation}
where $I_{\Delta \epsilon_{int}} (j',v') =1\  \forall \{j',v'\}$ such that $ \epsilon_{int}(j',v') \in  [\epsilon_{int}-\Delta \epsilon_{int}/2, \epsilon_{int}+\Delta \epsilon_{int} /2] $ and $0$ otherwise.

Fig~\ref{eint_erot_evib}  shows that an increment of $\Delta \epsilon_{int}$ results in higher vibrational energy relative to rotational energy (in the average sense) for bound molecules. However for quasi-bound molecules, an increment of internal energy adds more rotational energy relative to vibrational. Although elevated rotational energy does enhance dissociation, its effect is not as significant as for vibrational energy.
The final term (IV) in (Eq. \ref{prob_expression_main}) is justified by the analysis of QCT data shown in (Fig~\ref{prob_eint_QCT}), and captures the noticeable difference in dissociation probability trends for bound and quasi-bound molecules. Specifically, the functional form of term (IV) is guided by the QCT data plotted in Fig. \ref{prob_pop_QCT}, that shows a sharp peak for the internal energy of dissociating molecules centered at the dissociation energy, $\epsilon_d$, which is captured accurately using a constant value of the parameter $\delta$ (see Table \ref{parameters} in Sec. V (A)). This trend has been observed in QCT studies that assume Boltzmann distributions for the pre-collision distributions \cite{bender2015improved}.

\section{Integration of Cross-sections and Continuum Rate Expressions} \label{rate_constant_boltz_qss_nb}
Now that the state-specific dissociation cross-section model is formulated, the expression (Eq. \ref{prob_expression_main}) can be analytically integrated over a rovibrational distribution function, $f^{NB}(j,v)$, to derive a closed-form continuum-level dissociation rate expression using Eq. \ref{Eratemain}. The considerable challenge with such an approach is that if the cross-section model is highly accurate then it must be integrated over an equally-accurate distribution function. If the distribution function is inaccurate, then the overall rate expression will be inaccurate.   

This section first develops an accurate Boltzmann internal energy distribution function that is consistent with ab-intio results, after which, non-Boltzmann internal energy effects can be included. Subsections then derive continuum dissociation rate expressions corresponding to (i) Boltzmann internal energy distributions, (ii) Quasi-Steady-State (QSS) internal energy distribution functions that include depletion of high energy states, and (iii) generalized non-Boltzmann internal energy distribution functions that include both overpopulation of high-energy states during the rapid excitation phase and depletion during the QSS dissociation phase. All model versions are consistent and the generalized expression reduces, analytically, to the QSS and Boltzmann expressions under corresponding assumptions. Note that, for conciseness, some of the derivation steps are omitted
from the main article text and are included in the Appendix.

\subsection{Internal energy distribution function framework}
Constructing an analytical expression for the internal energy distributions, that is consistent with ab-initio data, requires further approximations. First, a joint rovibrational distribution function, $f(\epsilon_{rot},\epsilon_{v})$, is required such that total internal energy does not exceed the physically-allowed maximum internal energy. A general approach is to use a separate vibrational temperature in the distribution, such that the moment with respect to $\epsilon_v$ results in the average vibrational energy of the system; that is, $\langle \epsilon_v \rangle =  \sum_v^{v_{\mx}} \sum_j^{j_{\mx}(v)} \epsilon_v(v)  f(\epsilon_{rot},\epsilon_{v})$. This distribution is written as:
\begin{equation}
\begin{split}
    f(\epsilon_{rot},\epsilon_{v}) = \cfrac{(2j+1) \exp\left[-\cfrac{\epsilon_{rot}(j)}{k_B T_{rot}} \right]\exp\left[-\cfrac{\epsilon_{v}(v)}{k_B T_{v}}  \right]}{\Zn(T_{rot},T_{v})} ~,
    \label{joint_abinitio_model}
\end{split}    
\end{equation}
where the partition function, $Z(T_{rot},T_{v})$, is given by
\begin{equation}
\begin{split}
\Zn(T_{rot},T_{v}) = \sum_{v=0}^{v=v_{\mx}} \sum_{j=0}^{\epsilon_d^{\text{max}} - \epsilon_v}   (2j+1)\exp\left[-\frac{\epsilon_{rot}(j)}{{k_B T_{rot}}} \right]\exp\left[-\frac{\epsilon_{v}(v)}{{k_B T_{v}}} \right]~,
\end{split}    
\end{equation}
such that an equilibrium distribution is recovered when $T_{rot} = T_v = T$. 

\begin{figure}
    \subfigure[]
  {
    \includegraphics[width=3.00in,trim={0.1cm 0.1cm 0.0cm 0.1cm},clip]{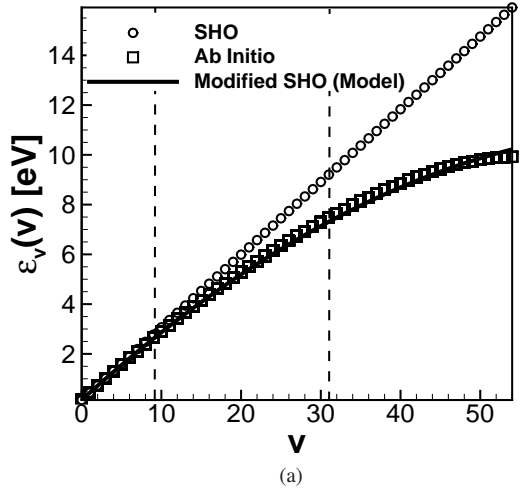}
     \label{model_ev}
   }  
       \subfigure[]
  {
    \includegraphics[width=3.0in]{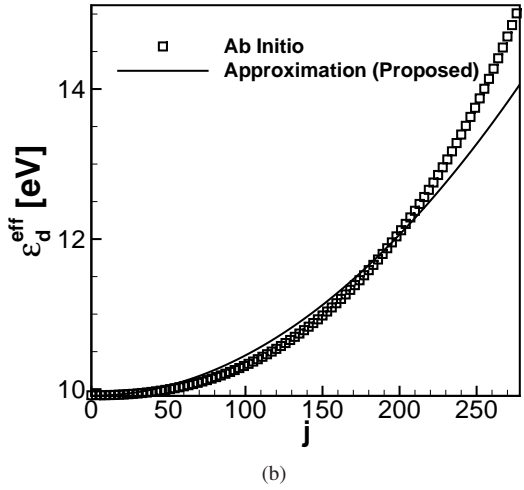}
     \label{centri_barrier}
   } 
   \caption{(a) Proposed approximation to $\epsilon(v)$ as function of vibrational quantum state. (b) Effective ($\epsilon_d^{\text{eff}}$) dissociation energy taking into account centrifugal barrier. }
   \label{abinitio-distros}
 \end{figure}

To complete this equation, one must express (i) vibrational energy as a function of vibrational quantum state, and (ii) rotational energy as a function of rotational quantum state. Rotational energy is approximated using the rigid rotor assumption: $\epsilon_{rot}(j)=\theta_{\text{rot}} k_B j(j+1)$. Vibrational energy is approximated using the simple harmonic oscillator (SHO) model, however, an important modification is required. Specifically, as shown in Fig.~\ref{model_ev}, SHO is a poor approximation of the vibrational energy (and therefore population) for moderate-to-high $v$ states compared to the results from the ab-initio diatomic nitrogen PES. Since the high-$v$ states are important, due to their strong coupling to dissociation, we use an extended SHO expression with multiple characteristic vibrational temperatures. In this article we use three characteristic temperatures, corresponding to the three energy brackets shown in Fig.~\ref{model_ev}. This introduces a summation (over three terms) into all subsequent rate expressions, however, the equations remain analytically integrable. One could use the anharmonic oscillator model, however, analytical integration would no longer be possible.

%
%

Furthermore, the value of $\theta_{CB}$, required to model the centrifugal barrier effect for $\epsilon^{\text{eff}}_{rot}$ and $\epsilon^{\text{eff}}_d$ in Eq. \ref{prob_expression_main}, is also determined by comparison with ab-intio data using the diatomic nitrogen PES. Specifically, the centrifugal barrier is incorporated in the effective dissociation energy in following manner, 
\begin{equation}
    \epsilon_d^{\text{eff}}=\epsilon_d+\theta_{CB} \epsilon_{rot}(j) \approx \epsilon_d+  j(j+1) \theta_{CB}\theta_{\text{rot}} k_B~,
\end{equation}
where $\theta_{CB}=$ is selected to match the \textit{ab initio} energy as shown in Fig.~\ref{centri_barrier}.

Finally, in Eq. \ref{joint_abinitio_model} we can replace $\epsilon_v (v)$ and $\epsilon_{rot}(j)$ using the modified SHO model (Fig. \ref{model_ev}) and the rigid rotor model, respectively. The expression corresponding to a Boltzmann distribution (at $T_{rot}, T_{v}$) contains a sum over $m$ terms ($m=3$ in our model), and can be derived as:
\begin{widetext}
\begin{equation}
\begin{split}
    f(j,v) = 
    (2j+1)\cfrac{\exp\left[-\cfrac{\theta_{rot}j(j+1)}{{T_{rot}}} \right]\exp\left[- \cfrac{E_{m^-}+ (v-m^-) \theta_v^mk_B }{{k_B  T_{v}}}  \right]}{\Zn(T_{rot},T_{v})}   \hspace{0.5 in} v \in [m^-,m^+)~,
    \label{boltz_distro}
\end{split}    
\end{equation}

where,
\begin{equation}
\begin{split}
    \Zn(T_{rot},T_{v}) =
     \frac{T_{rot}}{\theta_{rot}} \left \{ \gn \left(-\frac{1}{k_B T_v}\right)  - \exp\left[-\frac{\epsilon_d^{max}}{k_B T_{rot}}\right]  \gn \left(-\frac{1}{k_B T_v}+\frac{1}{k_B T_{rot}}\right)\right \}.
    \label{Ztvtr1}
\end{split}
\end{equation}
\end{widetext}
Here, $E_{m^-}$ is the energy of the $m^-$ quantum state and,
\begin{equation*}
   \gn(x)= \sum_m \gn_m(x)
\end{equation*}
\begin{equation*}
    \gn_m(x)=  \exp\left[x E_{m^-}\right]\frac{1-\exp[x (m^+-m^-) \theta_v^m\ k_B ]}{1-\exp[x\  \theta_v^m\ k_B ]} 
\end{equation*}
is the resulting generalized function that contains the summation information, noting that $g_m(0)\equiv \lim_{x \rightarrow 0} g_m(x)$. 
Upon inspection of Eq. \ref{boltz_distro}, one may realize that in the limit $\epsilon_d^{\text{max}} \rightarrow \infty$ and assuming a single characteristic vibrational temperature such that $\theta_v^m=\theta_v^{\text{SHO}}$ ($\forall m$), the partition function $\ Z(T_{rot},T_{v})$ reduces to the product of standard rigid rotor and SHO partition functions, consistent with equilibrium statistical mechanics. 
In order to evaluate the average vibrational energy of dissociated molecules ($\langle \epsilon_{v}^d \rangle$ in Eq. \ref{Edv_eqn}), the derivative of the function $g_m(x)$ is required: 
\begin{equation*}
   \gn(x)'= \sum_m \gn_m(x)'
\end{equation*}
\begin{equation*}
    \gn_m(x)'= \frac{\partial \gn_m}{\partial x} = \gn_m\frac{\partial \log \gn_m}{\partial x} 
\end{equation*}
\begin{widetext}
\begin{equation*}
\begin{split}
   \gn_m'(x)=\gn_m(x)\left\{ E_{m^-} - \frac{(m^+-m^-) \theta_v^m\ k_B \exp[x (m^+-m^-) \theta_v^m\ k_B ]}{1-\exp[x (m^+-m^-) \theta_v^m\ k_B ]} 
   + \frac{ \theta_v^m\ k_B \exp[x\  \theta_v^m\ k_B ]}{1-\exp[x\  \theta_v^m\ k_B ]} \right \} ~,
\end{split}
\end{equation*}
\end{widetext}
where $g_m'(0)\equiv \lim_{x \rightarrow 0} g_m'(x)$.

At this point, analytical expressions for both the dissociation cross-sections (Eq. \ref{prob_expression_main}) and the internal energy distribution functions (Eq. \ref{boltz_distro}) have been developed. In subsequent sections these expressions are analytically integrated (using Eq. \ref{Eratemain}) to produce continuum-level dissociation rate expressions.

\subsection{Boltzmann rate expression}\label{Boltz_Distro_Int}
Under the assumption that the gas internal energy evolves as a series of Boltzmann distributions, corresponding to $T_{rot}$ and $T_v$, the dissociation rate constant is analytically obtained by integrating the reaction probability (Eq.~\ref{prob_expression_main}) over the distributions in Eq.~\ref{boltz_distro} according to Eq.~\ref{Eratemain}:
 \begin{equation}
\begin{split}
   k(T, T_{rot},T_{v}) 
  = A T^{\eta} \exp\left[- \frac{\epsilon_d}{k_B T} \right]    *\left[\Hn(\epsilon_d,0,1) +  \Hn(\epsilon_d^{\mx},\epsilon_d,2) \right]
  \end{split}
  \label{Rate_Final_Boltz}
\end{equation}
\begin{equation*}
\eta =\alpha-\frac{1}{2}; \hspace{0.35in} A= \frac{1}{S}  \left(\frac{8 k_B }{\pi \mu_C}\right)^{1/2} \pi b_{\max}^2  C_1 \Gamma[1+\alpha] \left(\frac{k_B }{\epsilon_d}\right)^{\alpha-1}
\end{equation*}
\begin{equation}
\begin{split}
    \Hn(\epsilon_i,\epsilon_j,n) = \hspace{3.0in}\\ \frac{\exp[(-1)^{n-1} \delta ]}{\Zn(T_{rot},T_{v})} \frac{\exp\left[ \epsilon_i \zeta_{rot}\right]  \gn (\zeta_{vr}) - \exp\left[ \epsilon_j\zeta_{rot}\right] \gn (\zeta_{v}-\epsilon_j \zeta_{rot}/\epsilon_d )}{k_B \theta_{rot} \zeta_{rot}}  ;
    \label{HBoltz_gen}~,
\end{split}
\end{equation}
where,
\begin{equation*}
\begin{split}
\zeta_{rot}  = -\frac{1}{k_B T_{rot}}+\frac{1}{k_B T}
+\frac{\beta-\theta_{CB}+(-1)^n\delta}{\epsilon_d} -\frac{\theta_{CB}}{k_B T};\hspace{0.10 in}\\
\label{delta_defs}
\end{split}
\end{equation*}
\begin{equation*}
\begin{split}
\zeta_{v}  =  -\frac{1}{k_B T_v} +\frac{1}{k_B T}  +\frac{\gamma+(-1)^n\delta}{\epsilon_d}  \hspace{0.5in} \zeta_{vr} =\zeta_{v}-\zeta_{rot}
\label{delta_defs}
\end{split}
\end{equation*}

The result is a ``multi-temperature'' dissociation rate coefficient that consists of a modified Arrhenius rate term and a nonequilibrium term (denoted by the $H$ function) that acts to increase or decrease the Arrhenius rate based on the local internal energy content of the gas. Note that in contrast to the Park $\sqrt{TT_v}$ model, the internal energy dependence does not appear inside the Arrhenius term. Rather, the Arrhenius term is a function only of the translational temperature $T$, and the dependence on internal energy is entirely contained in the separate nonequilibrium term. More specifically, the pre-factor $A$ and the exponent $\eta$ depend on the parameter $\alpha$, as well as on species properties. Recall that $\alpha$ is the parameter that controls the translational energy dependence at the kinetic level (term I in Eq. \ref{prob_expression_main}). The internal energy dependence is contained in two terms, corresponding to contributions from both bound ($\Hn(\epsilon_d,0,1)$) and quasi-bound ($\Hn(\epsilon_d^{\mx},\epsilon_d,2)$) molecules. 

Since the equations are written in a compact generalized functional form, it is difficult to deduce physical interpretations directly. However, physical insight is evident after a few simplifications are introduced. Focusing only on the contribution from bound molecules, and assuming that rotational and vibrational energy are independent, the basic dependence of the nonequilibrium term is, 
\begin{equation}
    \Hn(\epsilon_d,0,1) \propto  \cfrac{\gn (\zeta_{v})}{-\zeta_{rot}}~.
\end{equation}
Here, $\gn (\zeta_{v})$ can be viewed as a partition function evaluated at an effective vibrational temperature of $T_F'$, where
\begin{equation}
   -\cfrac{1}{k_B T_F'} = -\frac{1}{k_B T_v}+\frac{1}{k_B T} +\frac{\gamma-\delta}{\epsilon_d}~.
   \label{TFp}
\end{equation}
In this case, increasing $T_v$ (and $\gamma$) will increase the rate constant due to a corresponding higher population of $v$-states. Interestingly, the resulting expression resembles the model of Marrone and Treanor \cite{marrone1963chemical}, where the effective temperature in their model is identical to the one given in Eq.~\ref{TFp}. Following the same simplifications, but now for quasi-bound molecules ($\Hn(\epsilon_d^{\mx},\epsilon_d,2)$), the effective temperature $T_{F'}$ according to our model is
\begin{equation}
   -\cfrac{1}{k_B T_F'} =  -\frac{1}{k_B T_v} + \frac{1}{k_B T_{rot}}
+\frac{\gamma-\beta+\theta_{CB}}{\epsilon_d} + \frac{\theta_{CB}}{k_B T}~.
\label{TFp-2}
\end{equation}

\subsection{Non-Boltzmann QSS rate expression}\label{QSS_Distro_Int}
It is well established, from ab-intio based calculations, that a significant portion of dissociation behind strong shock waves occurs in the quasi-steady-state (QSS) phase \cite{valentini2016dynamics,valentini2015direct2,valentini2014rovibrational,valentini2015N4,valentini2015N3,valentini2015direct,grover2019direct,grover2019jtht, panesi2014pre,macdonald2018construction_DMS,macdonald2018_QCT,birks1972,mccoy1977master,Singhpnas,Singhpnasplus}. The QSS phase is characterized by non-Boltzmann internal energy distribution functions, where the high energy states are depleted (relative to a Boltzmann distribution). This effect is caused by the increase in internal energy level populations due to translational-internal energy transfer being precisely balanced by the rapid removal of internal energy due to dissociation reactions. In order to obtain realistic dissociation rates, such non-Boltzmann QSS distributions must be accounted for. 

A simple analytical model for depleted internal energy distributions in the QSS phase has been previously constructed by the authors using a ``surprisal'' formulation \cite{Singhpnas,Singhpnasplus}. The non-Boltzmann QSS internal energy distributions (denoted here as $\hat{f}$) are given by:

\begin{equation}
    -\log\left[\frac{\hat{f}(i)}{f_0 (i;T)}\right]=\lambda_{0,i}+\lambda_{1,i} \frac{\langle \epsilon_t \rangle}{\epsilon_d} i~,
    \label{ned_model_v}
\end{equation}
where $i$ is either $j$ or $v$, $f_0 (i;T)$ is a Boltzmann distribution, $\lambda_{0,i}$ is the normalization constant, $\lambda_{1,i}$ is a constant parameter, which controls the extent of depletion and is based on the DMS simulation results \cite{Singhpnas}. The accuracy of this model will be discussed in the next section.
Using the approximations for \textit{ab initio} energies, the joint rovibrational distribution function corresponding to QSS conditions is:
\begin{widetext}
\begin{equation}
\begin{split}
    \hat{f}(j,v) = (2j+1)\cfrac{\exp\left[-\cfrac{\theta_{rot}j(j+1)}{{T_{rot}}} + \hat{\delta}_{rot} j(j+1) \right]\exp\left[- \cfrac{E_{m^-}+ (v-m^-) \theta_v^mk_B }{{k_B  T_{v}}}  + \hat{\delta}_{v} v\right]}{\hat{\Zn}(T_{rot},T_{v})}   \hspace{0.15 in} 
    \\
    v \in [m^-,m^+)
    \label{qss_distro}
\end{split}    
\end{equation}
\begin{equation}
\begin{split}
    \hat{\Zn}(T_{rot},T_{v}) = \cfrac{T_{rot}}{\theta_{rot} -\cfrac{ T_{rot}}{\theta_{rot}k_B}\hat{\delta}_{rot} }
    \left \{ \hat{\gn} \left(-\frac{1}{ k_B T_v}\right)  - \exp\left[-\cfrac{\epsilon_d^{\mx}}{k_B T_{rot}} +\epsilon_d^{\mx} \cfrac{\hat{\delta}_{rot}}{k_B \theta_{rot}} \right] \right. \\ \left. \times \hat{\gn} \left(-\frac{1}{ k_B T_v}+ \cfrac{1}{k_B T_{rot}} -\cfrac{\hat{\delta}_{rot}}{k_B \theta_{rot}}\right)\right \} ;
    \\
    \label{Ztvtr_qss}
\end{split}
\end{equation}
\begin{equation*}
    \hat{\delta}_v =  - \lambda_{1,v} \frac{  3 k_B T }{2\epsilon_d}  \hspace{0.25in}  \hat{\delta}_{rot}=  - \lambda_{1,j} \frac{  3 k_B T }{2\epsilon_d} \hspace{0.25in} ~,
\end{equation*}
where,
\begin{equation*}
  \hat{\gn}(x)= \sum_m \hat{\gn}_m(x)
\end{equation*}
\begin{equation*}
    \hat{\gn}_m(x)=  \exp\left[x E_{m^-}+ \hat{\delta}_v m^-\right]\frac{1-\exp[ (m^+-m^-) (x \theta_v^m\ k_B+ \hat{\delta}_v) ]}{1-\exp[x\  \theta_v^m\ k_B+ \hat{\delta}_v ]} 
\end{equation*}
\end{widetext}
Note that this expression is similar to the joint rovibrational Boltzmann distribution function (Eq. \ref{boltz_distro}), except for additional terms containing $\hat{\delta}_v$ and $\hat{\delta}_{rot}$, which account for depleted energy states due to dissociation. Also note that the above distribution (Eq.~\ref{qss_distro}) reduces to the Boltzmann distribution (Eq.~\ref{boltz_distro}) when the depletion terms vanish ($\hat{\delta}_v = 0$ and $\hat{\delta}_{rot} = 0$).

Now the reaction probability (Eq.~\ref{prob_expression_main}) is analytically integrated over the QSS distribution (Eq.~\ref{qss_distro}) according to Eq.~\ref{Eratemain} to obtain the dissociation rate constant corresponding to a depleted QSS internal energy distribution:
\begin{equation}
\begin{split}
   \hat{k}(T, T_{rot},T_{v}) 
  = A T^{\eta} \exp\left[- \frac{\epsilon_d}{k_B T} \right]    *\left[\hat{\Hn}(\epsilon_d,0,1) +  \hat{\Hn}(\epsilon_d^{\mx},\epsilon_d,2) \right]
  \end{split}
  \label{Rate_Final_QSS}
\end{equation}
\begin{equation*}
\eta =\alpha-\frac{1}{2}; \hspace{0.35in} A= \frac{1}{S}  \left(\frac{8 k_B }{\pi \mu_C}\right)^{1/2} \pi b_{\max}^2  C_1 \Gamma[1+\alpha] \left(\frac{k_B }{\epsilon_d}\right)^{\alpha-1}
\end{equation*}
\begin{equation}
\begin{split}
    \hat{\Hn}(\epsilon_i,\epsilon_j,n) = \hspace{3.0in} \\ \frac{\exp[(-1)^{n-1}\delta]}{ \hat{\Zn}(T_{rot},T_v)} \frac{\exp\left[ \epsilon_i \hat{\zeta}_{rot} \right]  \hat{\gn} (\hat{\zeta}_{vr}) - \exp\left[ \epsilon_j \hat{\zeta}_{rot} \right] \hat{\gn} (\zeta_{v}-\epsilon_j \hat{\zeta}_{rot}/\epsilon_d )}{k_B \theta_{rot} \hat{\zeta}_{rot}}~,
    \label{Ztvtr_rate_QSS}
\end{split}
\end{equation}
where,
\begin{equation*}
\begin{split}
\hat{\zeta}_{rot}  = -\frac{1}{k_B T_{rot}}+\frac{1}{k_B T}
+\frac{\beta-\theta_{CB}+(-1)^n\delta}{\epsilon_d} -\frac{\theta_{CB}}{k_B T}+\frac{\hat{\delta}_{rot} }{\theta_{rot} k_B};\hspace{0.10 in}\\
\label{deltarot_defs_QSS}
\end{split}
\end{equation*}
\begin{equation*}
\begin{split}
\zeta_{v}  =  -\frac{1}{k_B T_v} +\frac{1}{k_B T}  +\frac{\gamma+(-1)^n\delta}{\epsilon_d} ; \hspace{0.5in} \hat{\zeta}_{vr} = \zeta_{v}-\hat{\zeta}_{rot}
\label{deltav_defs_QSS}
\end{split}
\end{equation*}
By comparing Eq. \ref{Rate_Final_QSS} with Eq. \ref{Rate_Final_Boltz}, it is evident that effects due to the depletion of high-energy states in QSS appear as extra terms, but otherwise the expressions are identical. For example, $\hat{\delta}_{rot}$ now appears as an extra term in the $\hat{\zeta}_{rot}$ expression, which as outlined via Eqs. \ref{TFp} and \ref{TFp-2}, is analogous to changing the effective temperature of the partition function (i.e. effecting the population of states). The $\hat{\delta}_v$ term now appears inside the summation function $\hat{\gn}_m(x)$ simply because the vibrational energy partition function involves the modified SHO formulation shown previously in Fig. \ref{model_ev}. Both parameters ($\hat{\delta}_{rot}$ and $\hat{\delta}_v$) act to reduce the nonequilibrium $H$ factors and, therefore, reduce the dissociation rate coefficient compared to the Boltzmann rate coefficient expression (recovered by setting $\hat{\delta}_{rot}=\hat{\delta}_v=0$).

\subsection{Generalized non-Boltzmann rate expression}\label{Non_Boltz_Distro_Int}
It is also well established, from ab-intio based calculations, that a significant amount of post-shock dissociation can occur \emph{before} the gas reaches QSS \cite{sahai2019flow}, particularly for extreme flight conditions such as atmospheric re-entry. 
 \begin{figure}
  \centering 
   \includegraphics[width=3.0in,trim={0.15cm 0.15cm 0.15cm 0.15cm},clip]{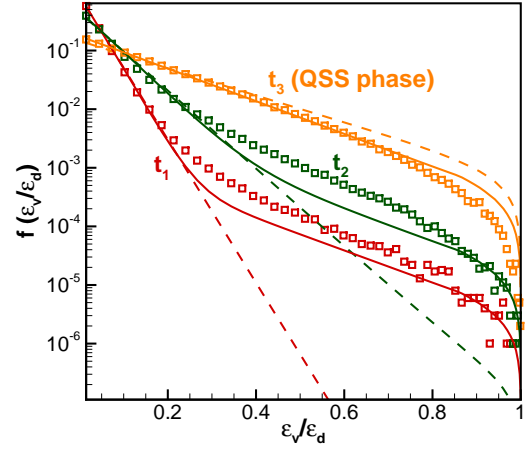}
    \label{Nitrogen_Distros_Isothermal}
 \caption{ 
 Vibrational energy distributions from the model (solid lines), DMS (symbols) and Boltzmann distribution (dashed line). DMS data is taken from Ref.~\cite{valentini2014rovibrational}. For more details about the model, please refer to \cite{Singhpnas} and \cite{Singhpnasplus}.}
  \label{Isothermal_Nitrogen_20K_DMS_Model}
\end{figure}

Figure \ref{Isothermal_Nitrogen_20K_DMS_Model} shows the evolution of vibrational energy distributions corresponding to rapid rovibrational excitation of nitrogen gas, starting from $T_v=$ 3,000 K and rising towards a fixed translational temperature of $T=$ 20,000 K. The data (symbols in Fig. \ref{Isothermal_Nitrogen_20K_DMS_Model}) come from recent Direct Molecular Simulation (DMS) calculations \cite{Singhpnas,Singhpnasplus}, and show that at early times during rapid rovibrational excitation the high-energy states are \emph{over-populated} compared to the Boltzmann distribution corresponding to the average vibrational energy (dashed lines in Fig. \ref{Isothermal_Nitrogen_20K_DMS_Model}). Since high-energy states are strongly coupled to dissociation, including the effects of overpopulation would increase the dissociation rate compared to the rate based on Boltzmann distributions (section 4.2). At later times, as the gas reaches QSS, Fig. \ref{Isothermal_Nitrogen_20K_DMS_Model} shows that the high-energy states are depleted relative to a Boltzmann distribution. This is the QSS distribution function described in section 4.3 (Eqs. \ref{ned_model_v} and \ref{qss_distro}).

Recently, the authors have developed a generalized model for non-Boltzmann distributions that captures both overpopulation and depletion \cite{Singhpnas}. In this section we derive the dissociation rate constant in its most general form including all non-Boltzmann effects. While full model details of the generalized non-Boltzmann distribution function ($f^{NB}(j,v)$) can be found in Ref. \cite{Singhpnasplus}, the final model equations are listed here for completeness:
\begin{equation}
     f^{NB}(j,v) = \cfrac{\tilde{f}(j,v;T_v,T_0)+\Lambda \hat{f}(j,v;T)}{1+\Lambda } ~,
     \label{fjvnb_overall}
\end{equation}
where $T_0$  is the reference temperature (based on the pre-shock state), the function $\hat{f}(j,v;T)$ is the QSS distribution in Eq. \ref{qss_distro}, and the function $\tilde{f}(j,v;T_v,T_0)$, is given by:
\begin{widetext}
  \begin{equation}
 \begin{split}
       \tilde{f}(j,v;T_v,T_0)  =  \hspace{4.0in} 
       \\
      \cfrac{(2j+1) \exp\left[-\cfrac{\Delta_{\epsilon} v}{k_B T_v}-\cfrac{\Delta_{\epsilon} v -\epsilon_v(v)}{k_B T_0}+ \hat{\delta}_{v} v\right]
       \exp\left[-\cfrac{\epsilon_{j}(j)}{{k_B T_{rot}}}  + \hat{\delta}_{rot} j(j+1) \right]}
       { \sum_v \sum_j^{v(j_{\mx})} (2j+1) \exp\left[-\cfrac{\Delta_{\epsilon} v}{k_B T_v}-\cfrac{\Delta_{\epsilon} v -\epsilon_v(v)}{k_B T_0}+ \hat{\delta}_{v} v\right]
       \exp\left[-\cfrac{\epsilon_{j}(j)}{{k_B T_{rot}}}  + \hat{\delta}_{rot} j(j+1) \right]}
       \label{sho_fnb}
        \end{split}
 \end{equation}
  \begin{equation*}
     \begin{split}
         \Delta_{\epsilon} = \epsilon_v(1)-\epsilon_v(0)
     \end{split}
 \end{equation*}
 \end{widetext}
Here, $T_v$ is temperature such that $ \sum_v \sum_j^{v(j_{\mx})} f(j,v) \epsilon_v(v)  = \langle \epsilon_v(v) \rangle$, and the parameter $\Lambda$ is analytically determined to ensure that $ \sum_v  f^{NB}(j,v)  \epsilon_v(v)  = \langle \epsilon_v(v) \rangle$. Specifically, $\Lambda$ is given by \cite{Singhpnasplus}:
\begin{equation}
    \Lambda = \cfrac{\langle \epsilon_v(v) \rangle - \langle \tilde{\epsilon_v} \rangle(T_v,T_0,T_{rot})}{ \langle \hat{\epsilon}_v \rangle (T_{rot},T_{v}) -\langle \epsilon_v(v) \rangle }
\end{equation}
\begin{equation}
    \langle \tilde{\epsilon_v} \rangle(T_v,T_0,T_{rot}) = \sum_{v=0}^{v_{\mx}} \sum_{j=0}^{v_{\mx}(j)} \epsilon_v(v) \tilde{f}(j,v;T_v,T_0,T_{rot})
\end{equation}

\begin{equation}
    \langle \hat{\epsilon}_v \rangle (T_{rot},T_{v}) = \sum_{v=0}^{v_{\mx}} \sum_{j=0}^{v_{\mx}(j)}  \epsilon_v(v)\hat{f}(j,v;T)
\end{equation}
where the expression for $\langle \tilde{\epsilon_v} \rangle(T_v,T_0,T_{rot})$ and $\langle \hat{\epsilon}_v \rangle (T_{rot},T_{v})$ are given in Eqs.~\ref{avgevtilde} and \ref{avgevhat} in  appendix \ref{lambdacalappend}.

As seen in Fig.~\ref{Isothermal_Nitrogen_20K_DMS_Model}, this simple analytical model quantitatively captures the evolution of non-Boltzmann vibrational energy distributions calculated by DMS. More comparisons for both nitrogen and oxygen gas under a range of conditions are contained in Ref. \cite{Singhpnasplus}. Following the same procedure as in sections 4.2 and 4.3, the approximations for \textit{ab initio} energies are substituted into Eq. \ref{fjvnb_overall} and the reaction probability (Eq.~\ref{prob_expression_main}) is analytically integrated according to Eq.~\ref{Eratemain} to obtain a generalized expression for the dissociation rate constant:
\begin{equation}
     k^{NB} = \cfrac{  \tilde{k}(T,T_{rot},T_v; T_0) +\Lambda \hat{k}(T,T_{rot},T) }  {1 + \Lambda } ~,
     \label{rate_full_nb}
\end{equation}
where, $\hat{k}(T, T_{rot},T_{v})$ was given in Eq. \ref{Rate_Final_QSS}, and $\tilde{k}(T,T_{rot},T_v; T_0)$ has a similar form, but based on the distribution in Eq.~\ref{sho_fnb}. For brevity, the equation for $\tilde{k}(T,T_{rot},T_v; T_0)$ is listed in the Appendix. Note, however, that the generalized non-Boltzmann rate coefficient expression (Eq. \ref{rate_full_nb}) has two contributing terms controlled by the value of $\Lambda$. In one limit ($T >> T_{rot}\approx T_{v}\approx T_0$, and thus $\Lambda=0$) the rate coefficient corresponds to the Boltzmann assumption at $T_0$ (post-shock frozen condition), while in the other limit ($T\approx T_{rot}\approx T_{v} >> T_0$, and thus $\Lambda^{-1}=0$) the rate coefficient corresponds to the QSS assumption. In between these limits both terms contribute and correspond to a non-Boltzmann distribution which may have overpopulation and/or depletion effects. Please refer to Ref.~\cite{Singhpnas} for complete details.

\section{Average vibrational energy of dissociated molecules}\label{edv_sec}
In addition to the dissociation rate coefficient, the other quantity required in continuum-level dissociation models is the average internal energy of dissociating molecules, 
$\langle \epsilon_v^d  \rangle$ (refer to Eq. \ref{LandauTeller_modified} and related discussion in section 2). Physically, $\langle \epsilon_v^d  \rangle$ is directly related to the state-resolved cross-section model and the non-Boltzmann internal energy distribution as quantified in Eq.~\ref{Edv_eqn}. 

\begin{figure}[ht]
  \centering
    \includegraphics[width=3.00in,trim={0.1cm 0.1cm 0.1cm 0.1cm},clip]{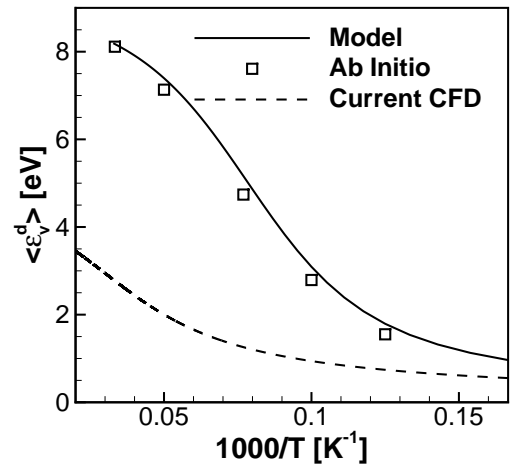}
   \label{prob_QCT_v}
   \caption{Average energy of dissociating molecules, current CFD uses $ \langle \epsilon_v^d \rangle  = \langle \epsilon_v \rangle$. \textit{Ab initio} data is taken from the QCT calculations presented by Chaudhry \textit{et al.} \cite{chaudhry2018qct}. }
   \label{edv_park_QCT}
 \end{figure}

The value of $\langle \epsilon_v^d \rangle$ recommended by Park\cite{park1989assessment} has no explicit dependence on cross-sections and exhibits numerical singularities \cite{candler2019rate}. The model for $\langle \epsilon_v^d \rangle$ currently used in CFD calculations, combined with the Park models for vibrational excitation and dissociation, is shown in Fig.~\ref{edv_park_QCT}, where significant discrepancy with ab-intio results is evident (for instance see Fig. 5 in the article by  Chaudhry \textit{et al.}\cite{chaudhry2018qct}).
In the following subsections, the analytical expression for $\langle \epsilon_v^d \rangle$ is derived using Eq.~\ref{Edv_eqn}, corresponding to the new dissociation model for Boltzmann, QSS, and generalized Boltzmann internal energy scenarios.

\subsection{Boltzmann internal energy distribution function formulation}
Employing Eq.~\ref{Edv_eqn}, assuming that gas evolves series of Boltzmann distributions, one can obtain an expression for 
average energy of dissociating molecules, $\langle \epsilon_v^d \rangle(T,T_{rot},T_{v})$ as:
\begin{equation}
    \langle \epsilon_v^d \rangle(T,T_{rot},T_{v}) =  \frac{\Phi(\epsilon_d,0,1)+\Phi(\epsilon_d^{\mx},\epsilon_d,2)}{\Hn(\epsilon_d,0,1)+\Hn(\epsilon_d^{\mx},\epsilon_d,2)}
    \label{edv_gen_sum_Boltz}
\end{equation}
where
\begin{equation*}
\begin{split}
    \Phi(\epsilon_i,\epsilon_j,n) = \hspace{3.0in} \\ \cfrac{\exp[(-1)^{n-1}\delta]}{Z(T_{rot},T_v)} \frac{\exp\left[ \epsilon_i \zeta_{rot}\right]  \gn' (k_B \theta_{rot}\zeta_{vr}) - \exp\left[ \epsilon_j \zeta_{rot}\right] \gn' (\zeta_{v}-\epsilon_j \zeta_{rot}/\epsilon_d )}{\zeta_{rot}} ;
    \label{Hnb_Boltz}
\end{split}
\end{equation*}
\begin{equation*}
\begin{split}
    \Hn(\epsilon_i,\epsilon_j,n) = \hspace{3.0in} \\   \cfrac{\exp[(-1)^{n-1}\delta]}{Z(T_{rot},T_v)}  \frac{\exp\left[ \epsilon_i \zeta_{rot}\right]  \gn (\zeta_{vr}) - \exp\left[ \epsilon_j\zeta_{rot}\right] \gn (\zeta_{v}-\epsilon_j \zeta_{rot}/\epsilon_d )}{k_B \theta_{rot}\zeta_{rot}};
    \label{Ztvtr_Boltz}
\end{split}
\end{equation*}
For the sake of brevity, we have presented the expressions for the average energy of dissociating molecules in the QSS formulation ($\langle \hat{\epsilon}^{d}_v\rangle(T,T_{rot},T_{v})$) and generalized non-Boltzmann case ($\langle \epsilon^{d}_v \rangle ^{NB}(T,T_{rot},T_{v})$) in the Eqs.~\ref{avg_edv_QSS} and \ref{avg_edv_NB} in the appendix \ref{avgedvappend}, respectively.

\section{Model Verification} \label{Results_and_Discussion}

In summary, the general non-Boltzmann version of the model (for $k$ and $\langle\epsilon_v^d \rangle$) is given by Eqs.~\ref{rate_full_nb} and \ref{avg_edv_NB}, which can be used in continuum CFD calculations as a two-temperature (or three-temperature) model. Recall that the model is actually formulated at the kinetic level, where the two required models included $p (d\ |\ \epsilon_{rel},\epsilon_{v},\epsilon_{rot})$ and $f^{NB}(j,v)$. All subsequent equations were analytically derived from these two expressions. Therefore, the generalized model includes dissociation effects due to translational, rotational, and vibrational energy, as well as quasi-bound effects, anharmonicity in the diatomic potential, and both overpopulation and depletion of high-energy states. The model parameters for nitrogen are listed in Table \ref{parameters}, and are the same parameters used to calculate all of the model results plotted in this article. The generalized model can be simplified to a model version that assumes the gas evolves as a series of QSS distributions by replacing $\Lambda \rightarrow \infty$, and simplified further to the assumption of Boltzmann internal energy distributions by substituting $\delta_{rot} = 0$ and $\delta_{vib} =0$. In this section, we verify model results against previous QCT results and compare against experimental data.

\begin{table}
\centering 
\caption{\label{tab:table-name} Constants for approximation of \textit{ab initio} energies and parameters required in the model.}
\begin{tabular}{ | m{13em} | m{4.2cm}| }
\hline
Vibrational energy  & $\theta_v^{I}=3390$ K for $\vv \in [0,9) $,\\ 
(SHO) &  $\theta_v^{II}=0.75\  \theta_v^{I}$  for $\vv \in [9,31) $,  \\ 
 &  $\theta_v^{III}=0.45\  \theta_v^{I}$  for $\vv \in [31,55) $  \\ 
\hline
Rotational energy (Rigid Rotor) & $\theta_{rot} = 2.3$  K \\ 
\hline
Centrifugal barrier & $\theta_{CB} = 0.27$  \\ 
\hline
Diatomic energies & $\epsilon_d = 9.91 \text{eV} , \epsilon_d^{\max}=14.5 \text{eV}$  \\ 
\hline
Reaction probability & $C_{1} = 8.67\times 10^{-5}, \ \alpha =1.04 $, 
\\
                      &  $ \beta=5.91, \ \gamma=3.49,\ \delta = 1.20$  
\\ 
\hline
Non-Boltzmann distributions \cite{Singhpnas} & $\lambda_{1,v}=0.080, \lambda_{1,j}=4.33\times 10^{-5} $  \\ 
\hline
\end{tabular}
\label{parameters}
\end{table}

   \begin{figure}
  \subfigure[Rate Constant]
  {
    \includegraphics[width=3.0in]{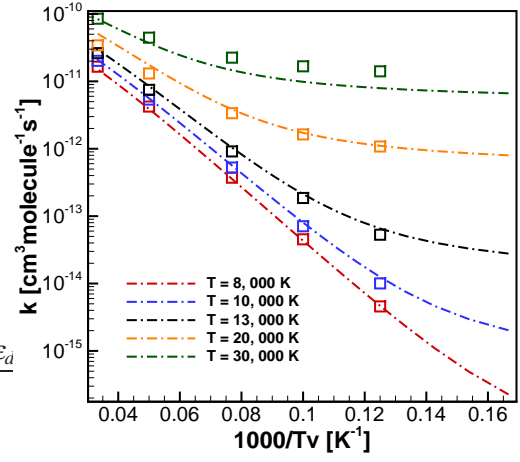}
   \label{Rate_QCT_model}
   }  
      \subfigure[$\langle \epsilon_v^d \rangle$ (Thermal Non-Equilibrium)]
  {
    \includegraphics[width=3.0in]{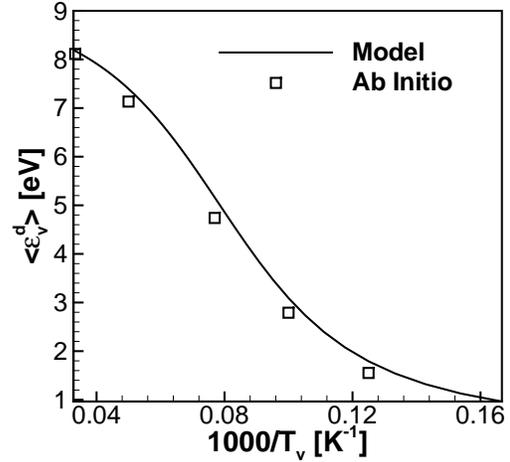}
\label{Edv_thermal_QCT}
   } 
        \subfigure[$\langle \epsilon_v^d \rangle$ (Thermal Equilibrium)]
  {
    \centering \includegraphics[width=3.0in]{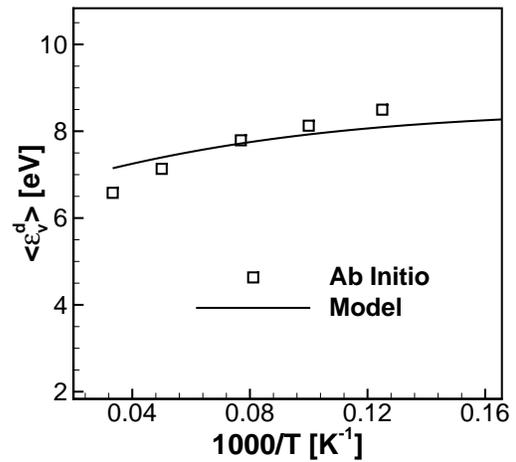}
   \label{Edv_nonthermal_QCT}
   }  
   \caption{(a) Reaction rate constants from QCT calculations \cite{bender2015improved} (shown with symbols) and the model (dashed line). Rotational and relative translation energies have Boltzmann and Maxwell-Boltzmann distributions at temperature $T$ respectively. Vibrational energy also has Boltzmann distribution at $T_v$, a quantity analogous to temperature. (b) Average energy of the dissociating molecules using the model (lines) and QCT \cite{bender2015improved} (symbols) for $T=T_{rot}=20,000$ K (c)(b) Average energy of the dissociating molecules using the model (lines) and QCT \cite{bender2015improved} (symbols) for $T=T_v=T_{rot}$}
   \label{QCT_Comparison}
 \end{figure}

Bender \textit{et.al.} \cite{bender2015improved} performed a QCT study of nitrogen dissociation where they considered translational-rotational equilibrium (based on temperature $T$) and Boltzmann distributions for vibrational energy based on a separate vibrational temperature ($T_v$). The internal state populations were sampled directly from the quantized $(j,v)$ states of the diatomic PES. We compare our model results (Eq.~\ref{Rate_Final_Boltz}) to these QCT results in Figure \ref{Rate_QCT_model} for the dissociation rate coefficient under vibrational nonequilibrium conditions ($T_v \neq T$). Specifically, our model results are plotted assuming Boltzmann distributions to be consistent with the results of Bender \textit{et al.}. The deviation between model and QCT results is seen to be less than $11 \%$ for the thermal equilibrium cases, less than $17\% $ for nonequilibrium cases where $T<13,0000$K, and a maximum difference of $50\%$ is found for the most extreme case, where $T=30,0000$K and $T_v=8000$K. 

Bender \textit{et.al.} \cite{bender2015improved}  also computed the average energy of dissociating molecules for the same equilibrium and thermal-nonequilibrium cases. The model results, again assuming Boltzmann distributions ($\langle \epsilon_v^d \rangle$ given in Eq.~\ref{edv_gen_sum_Boltz}), are compared to the QCT results in Figs.~\ref{Edv_thermal_QCT} and \ref{Edv_nonthermal_QCT}. Under thermal equilibrium conditions, it can be seen that the deviation between model and QCT results is less than $6\%$, and for nonequilibrium conditions ($T\neq T_v$), the agreement is even closer. 

It is important to note that, although the proposed model was formulated using QCT data from Bender \textit{et.al.} \cite{bender2015improved}, this was done at the kinetic level (Fig. \ref{prob_QCT_trans}), and obtaining close agreement at the rate level (Fig. \ref{QCT_Comparison}) is not trivial. Such close agreement requires an accurate model of the Boltzmann distribution including anharmonicity of the diatomic PES (the modified SHO formulation) and the use of a joint rovibrational distribution function, while also requiring an accurate model for the probability of dissociation due to each of translational, rotational, and vibrational energy, as well as the contribution from quasi-bound molecules. If any one of these effects is missing from the model, significant discrepancy results. Indeed, since there are no fitting parameters at the continuum-level of the model, this requires all kinetic aspects of the model to be included and modeled accurately. 

\begin{figure}
\center
\includegraphics[width=3.2in,trim={0.1cm 0.1cm 0.1cm 0.1cm},clip]{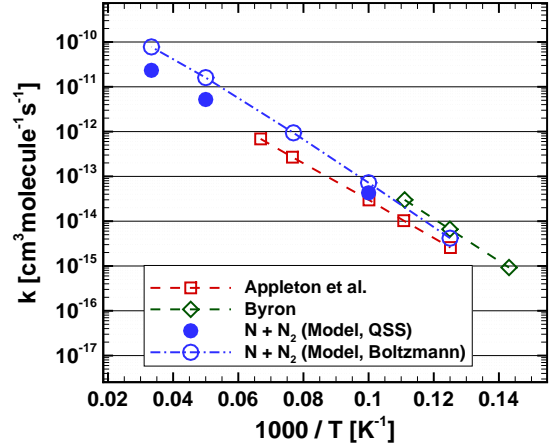}
       \caption{Comparison of rates from the present model for QSS phase, experiments \cite{appleton1968shock,byron1966shock}, QCT results from Bender \textit{et al.} \cite{bender2015improved} and DMS results from Valentini \textit{et al.} \cite{valentini2016dynamics}}
      \label{expt_dms_qss}
 \end{figure}

In Fig. \ref{expt_dms_qss} dissociation rates predicted by the model are compared to rates inferred from experimental shock-tube measurements by Appelton \textit{et al.} \cite{appelton} and Byron \cite{byron1966shock}. It is expected that the experimental flow conditions, under which the rate data was inferred, correspond to the QSS dissociation phase downstream of the shock wave. The model results are plotted for both Boltzmann internal energy distributions and for QSS internal energy distributions, and the dissociation rate constants are seen to be approximately 2-3 times lower under the QSS assumption. This trend has been reported a number of times in the literature in both master-equation studies \cite{macdonald2018_QCT} and direct molecular simulation (DMS) studies \cite{valentini2016dynamics} of nitrogen dissociation, and is reproduced by the new model. The disagreement (uncertainty) between various experimental results for nitrogen is approximately one order of magnitude, making model validation challenging. However, the model results lie within the variation of experimental results.

\begin{figure}
\centering 
\subfigure[Overpopulation Effect]
  {
    \includegraphics[width=3.2in]{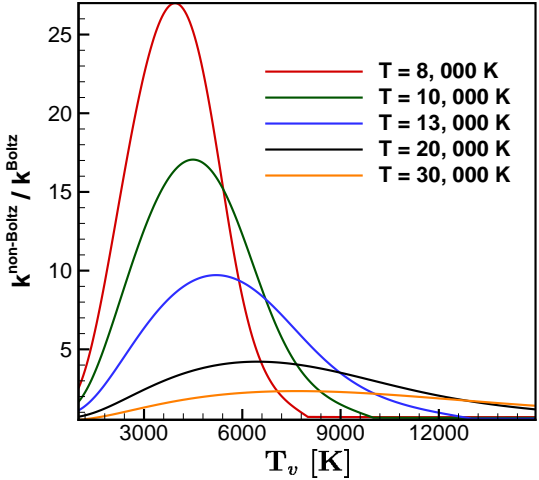}
   \label{overpopulation_transient_ratio}
   
   }  
   \subfigure[Depletion Effect]
 {
    \includegraphics[width=3.2in]{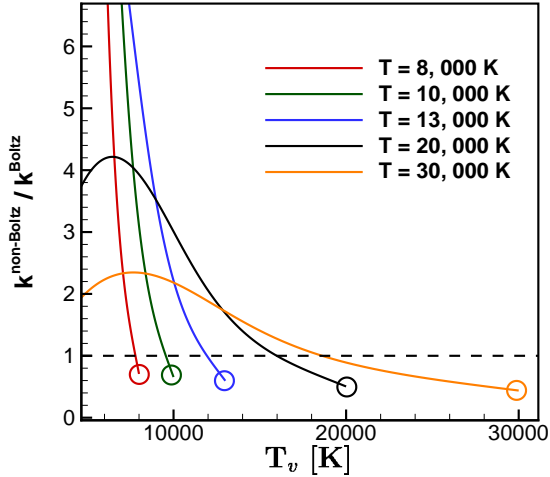}
    \label{depletion_transient_Ratio}
 }  
   \caption{Ratio ($k^{NB}/k$) of non equilibrium reaction rate constant to the rate constant at $T_v$ with Boltzmann distribution in transient phase. The final states denoted by circles (in Fig.~(b)) correspond to $\langle \epsilon_v \rangle  = \langle \epsilon_v \rangle^*$. $T_t=T_{rot}$ and the reference temperature $T_0 =300$ K is used. }
   \label{knb_kb_ratio}
\end{figure}

One of the primary goals of this work is to analyze the coupling of nonequilibrium effects on dissociation. Since the new model is analytically derived from the kinetic level, separate terms appear in the continuum expressions that include the various non-equilibrium effects of interest. This is advantageous, since the effect of each term can be quantitatively studied by comparing the magnitude of various terms or by switching terms of interest on or off. A detailed study of the importance of each nonequilibrium effect is contained in a subsequent article \cite{SinghCFDII}. As an example, the influence of overpopulation and depletion effects is shown in Fig. \ref{knb_kb_ratio}, where the ratio of the general nonequilibrium rate constant to the equilibrium rate constant is plotted as a function of vibrational temperature at various translational-rotational temperatures. 
In Fig. ~\ref{knb_kb_ratio}, it is evident that overpopulation of high-vibrational levels, in the rapid excitation phase ($T >> T_v$) increases the rate of dissociation (see Fig.~\ref{overpopulation_transient_ratio}). Clearly, overpopulation effects are more prominent at low translational temperature, raising the dissociation rate coefficient by as much as 27x compared to rate determined using the average vibrational energy (Boltzmann assumption). This can be explained from Fig.~\ref{Rate_QCT_model}, where rate constants depend on vibrational temperature, and consequently on the population of high v-states, more strongly at lower translational temperature.

When the gas reaches higher vibrational temperatures, corresponding to the QSS dissociating phase, as shown in Fig.~\ref{depletion_transient_Ratio}, the non-Boltzmann rates are lower due to the depletion of high vibrational energy states. The reduction in the dissociation rate coefficient under such QSS conditions ($T \approx T_v$) is rather constant across a wide range of translational temperatures, a result also found in master-equation and DMS studies \cite{macdonald2018_QCT,grover2019direct,valentini2016dynamics}.
It should be noted, however, that the extent of depletion and consequent reduction in the dissociation rate will be more prominent in an actual physical ensemble of gas, since the true QSS state has $\langle \epsilon_v \rangle  < \langle \epsilon_v \rangle^*$, which is not the same as the equilibrium condition $\langle \epsilon_v \rangle = \langle \epsilon_v \rangle^*$, plotted in Fig.~\ref{depletion_transient_Ratio}. Finally, although the effects of overpopulation appear dramatic (Fig.~\ref{overpopulation_transient_ratio}), the Boltzmann rate coefficient ($k^{Boltz}$) is low, since the average vibrational energy (represented by $T_v$) is low. Therefore, although overpopulation dramatically increases the dissociation rate relative to the Boltzmann assumption, it is not clear if this will noticeably affect the overall dissociation trend. A further study has therefore been performed where the new model is compared to DMS calculations and the relative importance of each nonequilibrium term is analyzed compared to the main rovibrational excitation and dissociation trends of importance to hypersonic CFD modeling (Ref.~\cite{SinghCFDII}).


\section{Conclusions}

We propose a cross-section model for dissociation with a simple functional form that captures recent \textit{ab initio} data  accurately. The model can be used in DSMC instead of large state-resolved databases. The model is analytically integrated over Boltzmann distributions and non-Boltzmann distributions to a derive general non-equilibrium dissociation model that can be used in large-scale CFD simulations. The model captures all key physics such as dependence on translational energy, rotational energy, vibrational energy, internal energy, centrifugal barrier, non-Boltzmann effects including overpopulation and depletion of high energy states, and species dependent parameters. The model is shown to reproduce the rates from QCT corresponding to Boltzmann distributions of internal energy states and for QSS distributions predicted by DMS. The QSS rates obtained from the model are compared with available experimental data. The effects of overpopulation inside shock waves and in the region immediately behind shock waves are also captured by the model. Further comparisons of the model are carried out in another article and simplifications to the model are proposed.    
 
\section*{Acknowledgement}
This work was supported by Air Force Office of Sci-
entific Research Grants  FA9550-16-1-0161 and  FA9550-19-1-0219 and was also partially supported
by Air Force Research Laboratory Grant FA9453-17-2-0081. Narendra Singh was partially supported by Doctoral Dissertation Fellowship at the University of Minnesota. Furthermore, discussions with Dr. R. Chaudhry, Dr. P. Valentini and Prof. G Candler are greatly acknowledged. 

\vspace{3.0in}
\appendix
\begin{widetext}
\section*{Appendix}
 \section{Analytical expression for the non-Boltzmann rate constant}
Denoting $\mathscr{F}$ as the internal energy distribution function and the corresponding rate constant as $\mathscr{K}(T, T_{rot},T_v)$, using Eq.~\ref{Eratemain} and Eq.~\ref{prob_expression_main}

 \begin{equation*}
\begin{split}
  \mathscr{K}(T, T_{rot},T_v) =  \hspace{5.0in}
     \end{split}
  \label{Rate_Derive1}
\end{equation*}
 \begin{equation}
\begin{split}
  \frac{1}{S} \left(\frac{8 k_B T}{\pi \mu_C}\right)^{1/2} \sum _{v=0}^{v_{\max}}  \sum _{j=0}^{j_{\max}} \int _0^\infty p(d|\epsilon_{rel}, \epsilon_{rot}, \epsilon_{v})   \pi b_{\max}^2  \left(\frac{\epsilon_{rel}}{k_B T}\right) \exp\left[- \frac{\epsilon_{rel}}{k_B T} \right] d \left(\frac{\epsilon_{rel}}{k_B T}\right) \mathscr{F}(\epsilon_{rot},\epsilon_{v})
  \\
  =\frac{1}{S} \left(\frac{8 k_B T}{\pi \mu_C}\right)^{1/2} \pi b_{\max}^2  C_1 \sum _{v=0}^{v_{\max}}  \sum _{j=0}^{j_{\max}}\exp\left[\beta \frac{\epsilon_{rot}^{\text{eff}}}{\epsilon_d}\right] \exp\left[\gamma \frac{\epsilon_{v}}{\epsilon_d}\right] \exp\left[\delta \frac{|\epsilon_{int}-\epsilon_d|}{\epsilon_d}\right] \mathscr{F}(\epsilon_{rot},\epsilon_{v})\\
  \times \int _{\epsilon_d^{\ef}-\epsilon_{int}}^\infty  \left[\frac{(\epsilon_{rel}+\epsilon_{int}-\epsilon_d^{\ef})^{\alpha}}{\epsilon_{rel}}\right] 
 \left(\frac{1}{\epsilon_d}\right)^{\alpha-1}    \left(\frac{\epsilon_{rel}}{k_B T}\right) \exp\left[- \frac{\epsilon_{rel}}{k_B T} \right] d \left(\frac{\epsilon_{rel}}{k_B T}\right) 
  \end{split}
  \label{Rate_Derive2}
\end{equation}
With substitution of $(\epsilon_{rel}+\epsilon_{int}-\epsilon_d^{\ef})/(k_B T) = x$, one can reduce the above expression to the following:
 \begin{equation}
\begin{split}
    \mathscr{K}(T, T_{rot},T_v)  =\frac{1}{S} \left(\frac{8 k_B T}{\pi \mu_C}\right)^{1/2} \pi b_{\max}^2  C_1 \exp\left[- \frac{\epsilon_d}{k_B T} \right]   \left(\frac{k_B T}{\epsilon_d}\right)^{\alpha-1} \int _{0}^\infty  x^{\alpha}
   \exp\left(- x  \right) d x \times \\
   \left\{ \sum _{v=0}^{v_{\max}}  \sum _{j=0}^{j_{\max}} \exp\left[\beta \frac{\epsilon_{rot}^{\text{eff}}}{\epsilon_d}\right] \exp\left[\gamma \frac{\epsilon_{v}}{\epsilon_d}\right] \exp\left[\delta \frac{|\epsilon_{int}-\epsilon_d|}{\epsilon_d}\right]
  \exp\left[\frac{\epsilon_{int}}{k_B T} \right]  \exp\left[-\frac{\theta_{CB}\epsilon_{rot}}{k_B T} \right]
  \mathscr{F}(\epsilon_{rot},\epsilon_{v}) \right\}
   \end{split}
  \label{Rate_Derive3}
\end{equation}
Solving the integral above ($\alpha>-1$ is required for convergence, which is the case):
 \begin{equation}
\begin{split}
    \mathscr{K}(T, T_{rot},T_v)  =\frac{1}{S} \left(\frac{8 k_B T}{\pi \mu_C}\right)^{1/2} \pi b_{\max}^2  C_1 \exp\left[- \frac{\epsilon_d}{k_B T} \right]   \left(\frac{k_B T}{\epsilon_d}\right)^{\alpha-1} \Gamma[1+\alpha] \times \\
   \left\{ \sum _{v=0}^{v_{\max}}  \sum _{j=0}^{j_{\max}} \exp\left[\beta \frac{\epsilon_{rot}^{\text{eff}}}{\epsilon_d}\right] \exp\left[\gamma \frac{\epsilon_{v}}{\epsilon_d}\right] \exp\left[\delta \frac{|\epsilon_{int}-\epsilon_d|}{\epsilon_d}\right]
  \exp\left[\frac{\epsilon_{int}}{k_B T} \right]
  \exp\left[-\frac{\theta_{CB}\epsilon_{rot}}{k_B T} \right]
  \mathscr{F}(\epsilon_{rot},\epsilon_{v}) \right\}
   \end{split}
  \label{Rate_Derive4}
\end{equation}
where the quantity in curly brackets can be expressed as contribution from bound and quasi-bound molecules. In the above Eq.~\ref{Rate_Derive4}, depending upon the expression of $\mathscr{F}(\epsilon_{rot},\epsilon_{v})  ( = f(\epsilon_{rot},\epsilon_{v}),\hat{f}(\epsilon_{rot},\epsilon_{v})$ or $f^{NB}(\epsilon_{rot},\epsilon_{v}))$ corresponding Boltzmannn ($\mathscr{K}(T, T_{rot},T_v) =k (T, T_{rot},T_v)$, QSS $\mathscr{K}(T, T_{rot},T_v) = \hat{k}(T, T_{rot},T_v)$  and non-Boltzmann rates ($\mathscr{K}(T, T_{rot},T_v) = k^{NB}(T, T_{rot},T_v))$  can be obtained. 

The expressions for $k (T, T_{rot},T_v)$ and $ \hat{k}(T, T_{rot},T_v)$ are derived in the main article. Here, we present full non-Boltzmann rates which includes $\tilde{k}(T, T_{rot},T_v)$ $f^{NB}(\epsilon_{rot},\epsilon_{v})$
 \begin{equation}
     k^{NB} = \cfrac{  \tilde{k}(T,T_{rot},T_v; T_0) +\Lambda \hat{k}(T,T_{rot},T) }  {1 + \Lambda } 
     \label{rate_full_nb_append}
\end{equation}
where,
 \begin{equation}
\begin{split}
    \tilde{k}(T,T_{rot},T_v; T_0)  =\frac{1}{S} \left(\frac{8 k_B T}{\pi \mu_C}\right)^{1/2} \pi b_{\max}^2  C_1 \exp\left[- \frac{\epsilon_d}{k_B T} \right]   \left(\frac{k_B T}{\epsilon_d}\right)^{\alpha-1} \Gamma[1+\alpha] \times \\
   \left\{ \sum _{v=0}^{v_{\max}}  \sum _{j=0}^{j_{\max}} \exp\left[\beta \frac{\epsilon_{rot}^{\text{eff}}}{\epsilon_d}\right] \exp\left[\gamma \frac{\epsilon_{v}}{\epsilon_d}\right] \exp\left[\delta \frac{|\epsilon_{int}-\epsilon_d|}{\epsilon_d}\right]
  \exp\left[\frac{\epsilon_{int}}{k_B T} \right]
  \exp\left[-\frac{\theta_{CB}\epsilon_{rot}}{k_B T} \right]
  \tilde{f}(j,v;T_v,T_0) \right\}
   \end{split}
  \label{Rate_Derive_tilde}
\end{equation}
The expression for $\hat{k}(T,T_{rot},T)$ is given by Eq.~\ref{Rate_Final_QSS} in the main manuscript. Before we perform the above summation (in Eq.~\ref{Rate_Derive_tilde}) we evaluate the partition function. 

\section{Quantities Related to $\tilde{f}(j,v;T_v,T_0)$}

\subsection{Partition Function for $\tilde{f}(j,v;T_v,T_0)$  }
The expression for the distribution function for $\tilde{f}(j,v;T_v,T_0)$ is,
  \begin{equation}
 \begin{split}
       \tilde{f}(j,v;T_v,T_0) = \cfrac{(2j+1) \exp\left[-\cfrac{\Delta_{\epsilon} v}{k_B T_v}-\cfrac{\Delta_{\epsilon} v -\epsilon_v(v)}{k_B T_0}+ \hat{\delta}_{v} v\right]
       \exp\left[-\cfrac{\epsilon_{j}(j)}{{k_B T_{rot}}}  + \hat{\delta}_{rot} j(j+1) \right]}
       { \sum_v \sum_j^{v(j_{\mx})}  (2j+1) \exp\left[-\cfrac{\Delta_{\epsilon} v}{k_B T_v}-\cfrac{\Delta_{\epsilon} v -\epsilon_v(v)}{k_B T_0}+ \hat{\delta}_{v} v\right]
       \exp\left[-\cfrac{\epsilon_{j}(j)}{{k_B T_{rot}}}  + \hat{\delta}_{rot} j(j+1) \right]}
       \label{sho_fnb_derive1}
        \end{split}
 \end{equation}
  \begin{equation*}
     \begin{split}
         \Delta_{\epsilon} = \epsilon_v(1)-\epsilon_v(0)
     \end{split}
 \end{equation*}
 where $T_v$ is temperature such that $ \sum_v \sum_j^{v(j_{\mx})} f(j,v) \epsilon_v(v)  = \langle \epsilon_v(v) \rangle$, and $\Lambda$ ensures that $ \sum_v  f^{NB}(j,v)  \epsilon_v(v)  = \langle \epsilon_v(v) \rangle$. Substituting the variation of $\epsilon_v(v)$ by modified SHO assumptions and $\epsilon_{rot}$ by rigid rotor assumptions, we can obtain an expression for $ \tilde{f}(j,v;T_v,T_0)$ as:
 \begin{equation*}
 \begin{split}
       \tilde{f}(j,v;T_v,T_0,T_{rot}) = \hspace{5.0 in}
\end{split}
\end{equation*}
  \begin{equation}
 \begin{split}
         \cfrac{(2j+1) \exp\left[-\cfrac{\Delta_{\epsilon} v}{k_B T_v}-\cfrac{\Delta_{\epsilon} v}{k_B T_0}+\cfrac{E_{m^-}+ (v-m^-) \theta_v^mk_B}{k_B T_0}+ \hat{\delta}_{v} v\right] 
       \exp\left[-\cfrac{\theta_{rot}j(j+1)}{{T_{rot}}}  + \hat{\delta}_{rot} j(j+1) \right]}
       { \tilde{Z} (T_v,T_0,T_{rot}) }
       \label{sho_fnb_derive}
        \end{split}
 \end{equation}
 where
  \begin{equation*}
     \begin{split}
         \Delta_{\epsilon} = \epsilon_v(1)-\epsilon_v(0) = k_B \theta_v^I
     \end{split}
 \end{equation*}
 and the required partition function $\tilde{Z} (T_v,T_0,T_{rot})$ is
   \begin{equation*}
     \begin{split}
         \tilde{Z} (T_v,T_0,T_{rot}) =  \sum_v^{v_{\mx}} \sum_j^{v(j_{\mx})}  (2j+1) \exp\left[-\cfrac{\Delta_{\epsilon} v}{k_B T_v}-\cfrac{\Delta_{\epsilon} v}{k_B T_0}+\cfrac{E_{m^-}+ (v-m^-) \theta_v^mk_B}{k_B T_0}+ \hat{\delta}_{v} v\right]   \exp\left[-\cfrac{\theta_{rot}j(j+1)}{{T_{rot}}}  + \hat{\delta}_{rot} j(j+1) \right]
     \end{split}
 \end{equation*}

 \begin{equation}
\begin{split}
    \tilde{\Zn}(T_v,T_0,T_{rot}) = \cfrac{T_{rot}}{\theta_{rot} -\cfrac{ T_{rot}}{\theta_{rot}k_B}\hat{\delta}_{rot} }
    \left \{ \tilde{\gn} \left(+\frac{1}{ k_B T_0}\right)  - \exp\left[-\cfrac{\epsilon_d^{\mx}}{k_B T_{rot}} +\epsilon_d^{\mx} \cfrac{\hat{\delta}_{rot}}{k_B \theta_{rot}} \right] \right.  \left. \times \tilde{\gn} \left(+\frac{1}{ k_B T_0}+ \cfrac{1}{k_B T_{rot}} -\cfrac{\hat{\delta}_{rot}}{k_B \theta_{rot}}\right)\right \} ;
    \label{Ztvtr_qss_derive}
\end{split}
\end{equation}
 
\end{widetext}

\begin{equation*}
    \hat{\delta}_v =  - \lambda_{1,v} \frac{  3 k_B T }{2\epsilon_d}  \hspace{0.25in}  \hat{\delta}_{rot}=  - \lambda_{1,j} \frac{  3 k_B T }{2\epsilon_d} \hspace{0.25in} 
\end{equation*}

where,
\begin{equation*}
  \tilde{\gn}(x)= \sum_m \tilde{\gn}_m(x)
\end{equation*}

\begin{equation*}
    \tilde{\gn}_m(x)=  \exp\left[x E_{m^-}+ \tilde{\delta}_v m^-\right]\frac{1-\exp[ (m^+-m^-) (x \theta_v^m\ k_B+ \tilde{\delta}_v) ]}{1-\exp[x\  \theta_v^m\ k_B+ \tilde{\delta}_v ]} 
\end{equation*}

\begin{equation*}
   \tilde{\delta}_v =  \hat{\delta}_v  -\cfrac{\Delta_{\epsilon} }{k_B T_v}-\cfrac{\Delta_{\epsilon} }{k_B T_0} \hspace{0.25in} 
\end{equation*}

where $\hat{\delta}_v$ and $\hat{\delta}_{rot}$ accounts for the depletion in the population due to dissociation.

\subsection{Rate Constant $ \tilde{k}(T,T_{rot},T_v; T_0)$}
With the expression for $\tilde{\Zn}(T_{rot},T_{v})$, we can now evaluate the sum in Eq.~\ref{Rate_Derive_tilde} to obtain $\tilde{k}(T,T_{rot},T_v; T_0)$ as follows:
\begin{equation}
\begin{split}
 \tilde{k}(T,T_{rot},T_v; T_0) 
  = A T^{\eta} \exp\left[- \frac{\epsilon_d}{k_B T} \right]    *\left[\tilde{\Hn}(\epsilon_d,0,1) +  \tilde{\Hn}(\epsilon_d^{\mx},\epsilon_d,2) \right]
  \end{split}
  \label{Rate_Final__derive}
\end{equation}

\begin{equation*}
\eta =\alpha-\frac{1}{2}; \hspace{0.35in} A= \frac{1}{S}  \left(\frac{8 k_B }{\pi \mu_C}\right)^{1/2} \pi b_{\max}^2  C_1 \Gamma[1+\alpha] \left(\frac{k_B }{\epsilon_d}\right)^{\alpha-1}
\end{equation*}

\begin{equation}
\begin{split}
    \tilde{\Hn}(x,y,n) = \hspace{3.0in} \\ \frac{\exp[(-1)^{n-1}\delta]}{ \tilde{\Zn}(T_v,T_0,T_{rot})} \frac{\exp\left[ x \hat{\zeta}_{rot} \right]  \tilde{\gn} (\hat{\zeta}_{vr}) - \exp\left[ y\hat{\zeta}_{rot} \right] \tilde{\gn} (\zeta_{v}-y\hat{\zeta}_{rot}/\epsilon_d )}{k_B \theta_{rot} \hat{\zeta}_{rot}}  ;
    \label{Ztvtr_rate__derive}
\end{split}
\end{equation}

where
\begin{equation*}
\begin{split}
\hat{\zeta}_{rot}  = -\frac{1}{k_B T_{rot}}+\frac{1}{k_B T}
+\frac{\beta-\theta_{CB}+(-1)^n\delta}{\epsilon_d} -\frac{\theta_{CB}}{k_B T}+\frac{\hat{\delta}_{rot} }{\theta_{rot} k_B};\hspace{0.10 in}\\
\label{deltarot_defs_QSS}
\end{split}
\end{equation*}
\begin{equation*}
\begin{split}
\zeta_{v}  =  +\frac{1}{k_B T_0} +\frac{1}{k_B T}  +\frac{\gamma+(-1)^n\delta}{\epsilon_d} ; \hspace{0.5in} \hat{\zeta}_{vr} = \zeta_{v}-\hat{\zeta}_{rot}
\label{deltav_defs_QSS}
\end{split}
\end{equation*}

\subsection{Expression for $\tilde{g}'(x)$}
The derivatives of $\tilde{\gn}$ can be expressed in the following manner:
\begin{equation*}
   \tilde{\gn}'(x)= \sum_m \tilde{\gn}'_m(x)
\end{equation*}
\begin{equation*}
    \tilde{\gn}'_m(x)= \frac{\partial \tilde{\gn}_m}{\partial x} = \tilde{\gn}_m\frac{\partial \log \tilde{\gn}_m}{\partial x} 
\end{equation*}

\begin{widetext}
\begin{equation*}
\begin{split}
   \tilde{\gn}'_m(x)=\tilde{\gn}_m(x)\left\{ E_{m^-} - \frac{(m^+-m^-)  \theta_v^m\ k_B \exp[ (m^+-m^-) ( x \theta_v^m\ k_B +\tilde{\delta}_v )]}{1-\exp[ (m^+-m^-)( x \theta_v^m\ k_B +\tilde{\delta}_v)]} \right. \\
   \left.
   + \frac{ \theta_v^m\ k_B \exp[x\  \theta_v^m\ k_B +\tilde{\delta}_v ]}{1-\exp[x\  \theta_v^m\ k_B +\tilde{\delta}_v  ]} \right \} 
\end{split}
\end{equation*}
\end{widetext}
\section{Average Energy of Dissociating Molecules}{\label{avgedvappend}}
\subsection{QSS Distribution Based Formulation}
In a manner similar to one used in the earlier subsection, for depleted distributions, the average energy of the dissociating molecules can be obtained:
\begin{equation}
    \langle \hat{\epsilon}^{d}_v\rangle(T,T_{rot},T_{v}) =  \frac{\hat{\Phi}(\epsilon_d,0,1)+\hat{\Phi}(\epsilon_d^{\mx},\epsilon_d,2)}{\hat{\Hn}(\epsilon_d,0,1)+\hat{\Hn}(\epsilon_d^{\mx},\epsilon_d,2)}
    \label{avg_edv_QSS}
\end{equation}
where
\begin{equation}
\begin{split}
    \hat{\Phi}(\epsilon_i,\epsilon_j,n) = \hspace{3.0in} \\
    \cfrac{\exp[(-1)^{n-1}\delta]}{\hat{\Zn}(T_{rot},T_{v})} \frac{\exp\left[ \epsilon_i \hat{\zeta}_{rot}\right]  \hat{\gn}' (\hat{\zeta}_{vr}) - \exp\left[ \epsilon_j \hat{\zeta}_{rot}\right] \hat{\gn}' (\zeta_{v}-\epsilon_j \hat{\zeta}_{rot}/\epsilon_d )}{k_B \theta_{rot}\hat{\zeta}_{rot}} ;
    \label{Hnb_QSS_11}
\end{split}
\end{equation}
\begin{equation}
\begin{split}
    \hat{\Hn}(\epsilon_i,\epsilon_j,n) =  \hspace{3.0in} \\  \cfrac{\exp[(-1)^{n-1}\delta]}{\hat{\Zn}(T_{rot},T_{v})}  \frac{\exp\left[ \epsilon_i \hat{\zeta}_{rot}\right]  \hat{\gn} (\hat{\zeta}_{vr}) - \exp\left[ \epsilon_j \hat{\zeta}_{rot}\right] \hat{\gn} (\zeta_{v}-\epsilon_j \zeta_{rot}/\epsilon_d )}{k_B \theta_{rot} \hat{\zeta}_{rot}} ;
    \label{Ztvtr_QSS_12}
\end{split}
\end{equation}
where the derivatives of $\hat{\gn}$ can be expressed in the following manner:
\begin{equation}
   \hat{\gn}'(x)= \sum_m \hat{\gn}'_m(x)
   \label{hatgderivativesum}
\end{equation}
\begin{equation}
    \hat{\gn}'_m(x)= \frac{\partial \hat{\gn}_m}{\partial x} = \hat{\gn}_m\frac{\partial \log \hat{\gn}_m}{\partial x} 
    \label{hatgderivative_m}
\end{equation}

\begin{widetext}
\begin{equation}
\begin{split}
   \hat{\gn}'_m(x)=\hat{\gn}_m(x)\left\{ E_{m^-} - \frac{(m^+-m^-)  \theta_v^m\ k_B \exp[ (m^+-m^-) ( x \theta_v^m\ k_B +\hat{\delta}_v )]}{1-\exp[ (m^+-m^-)( x \theta_v^m\ k_B +\hat{\delta}_v)]} \right. \\ \left.
   + \frac{ \theta_v^m\ k_B \exp[x\  \theta_v^m\ k_B +\hat{\delta}_v ]}{1-\exp[x\  \theta_v^m\ k_B +\hat{\delta}_v  ]} \right \} 
\end{split}
\label{hatgderivative}
\end{equation}
\end{widetext}

\subsection{General Non Boltzmann Distribution Formulation}
An approximation to the average energy of dissociated molecules, in the manner analogous to the rate constant, is calculated as:
\begin{equation}
   \langle \epsilon^{d}_v \rangle ^{NB}(T,T_{rot},T_{v}) =  \cfrac{\langle \tilde{\epsilon}^{d}_v \rangle(T,T_{rot},T_v; T_0) + \langle \hat{\epsilon}^{d}_v\rangle(T,T_{rot},T_{v}) \Lambda  k_r}{1+\Lambda  k_r}
    \label{avg_edv_NB}
\end{equation}
where 
\begin{equation}
    k_r =\cfrac{\hat{k}(T,T_{rot},T) }{\tilde{k}(T,T_{rot},T_v; T_0)}
\end{equation}
and where the expression for $\langle \hat{\epsilon}^{d}_v\rangle(T,T_{rot},T_{v}$ is from Eq. 36 and $\tilde{\epsilon}^{d}_v(T,T_{rot},T_v; T_0)$ is given as follows:

\begin{equation}
    \langle \tilde{\epsilon}^{d}_v\rangle(T,T_{rot},T_{v}) =  \frac{\tilde{\Phi}(\epsilon_d,0,1)+\tilde{\Phi}(\epsilon_d^{\mx},\epsilon_d,2)}{\tilde{\Hn}(\epsilon_d,0,1)+\tilde{\Hn}(\epsilon_d^{\mx},\epsilon_d,2)}
    \label{avg_edv_QSS_T}
\end{equation}
where
\begin{equation*}
\begin{split}
    \tilde{\Phi}(\epsilon_i,\epsilon_j,n) = \hspace{3.0in} \\ \cfrac{\exp[(-1)^{n-1}\delta]}{\tilde{\Zn}(T_v,T_0,T_{rot})} \frac{\exp\left[ \epsilon_i \hat{\zeta}_{rot}\right]  \tilde{\gn}' (\hat{\zeta}_{vr}) - \exp\left[ \epsilon_j \hat{\zeta}_{rot}\right] \tilde{\gn}' (\zeta_{v}-\epsilon_j \hat{\zeta}_{rot}/\epsilon_d )}{k_B \theta_{rot} \hat{\zeta}_{rot}} ;
    \label{Hnb_QSS_T}
\end{split}
\end{equation*}
\begin{equation*}
\begin{split}
    \tilde{\Hn}(\epsilon_i,\epsilon_j,n) =   \hspace{3.0in} \\ \cfrac{\exp[(-1)^{n-1}\delta]}{\tilde{\Zn}(T_v,T_0,T_{rot})}  \frac{\exp\left[ \epsilon_i \hat{\zeta}_{rot}\right]  \tilde{\gn} (\hat{\zeta}_{vr}) - \exp\left[ \epsilon_j \hat{\zeta}_{rot}\right] \tilde{\gn} (\zeta_{v}-\epsilon_j \zeta_{rot}/\epsilon_d )}{k_B \theta_{rot} \hat{\zeta}_{rot}} ;
    \label{Ztvtr_QSS_T}
\end{split}
\end{equation*}

\section{Calculation of the parameter $\Lambda$} \label{lambdacalappend}
The parameter $\Lambda$ requires two quantities $\langle \tilde{\epsilon_v} \rangle$ and $ \langle \hat{\epsilon}_v \rangle$, which are mathematically described as:
\begin{equation}
    \langle \tilde{\epsilon_v} \rangle(T_v,T_0,T_{rot}) = \sum_{v=0}^{v_{\mx}} \sum_{j=0}^{v_{\mx}(j)} \epsilon_v(v) \tilde{f}(j,v;T_v,T_0,T_{rot})
\end{equation}

\begin{equation}
    \langle \hat{\epsilon}_v \rangle (T_{rot},T_{v}) = \sum_{v=0}^{v_{\mx}} \sum_{j=0}^{v_{\mx}(j)}  \epsilon_v(v)\hat{f}(j,v;T)
\end{equation}

The above expressions can be evaluated as:

\begin{widetext}
\begin{equation}
\begin{split}
    \langle \tilde{\epsilon_v} \rangle (T_v,T_0,T_{rot}) = \cfrac{1}{\tilde{Z}(T_v,T_0,T_{rot})}\cfrac{T_{rot}}{\left[\theta_{rot} -\cfrac{ T_{rot}}{\theta_{rot}k_B}\hat{\delta}_{rot} \right]} \left \{ -\tilde{\gn} '\left(+\frac{1}{ k_B T_0}\right) \right. \\ \left. + \exp\left[-\cfrac{\epsilon_d^{\mx}}{k_B T_{rot}} +\epsilon_d^{\mx} \cfrac{\hat{\delta}_{rot}}{k_B \theta_{rot}} \right]  \times \tilde{\gn}' \left(+\frac{1}{ k_B T_0}+ \cfrac{1}{k_B T_{rot}} -\cfrac{\hat{\delta}_{rot}}{k_B \theta_{rot}}\right)\right \}
    \label{avgevtilde}
    \end{split}
\end{equation}

\begin{equation}
\begin{split}
    \langle \hat{\epsilon_v} \rangle (T_{rot},T_{v}) = \cfrac{1}{\hat{\Zn}(T,T)}\cfrac{T_{rot}}{\left[ \theta_{rot} -\cfrac{ T_{rot}}{ \theta_{rot}k_B}\hat{\delta}_{rot} \right] }
    \left \{- \hat{\gn}' \left(-\frac{1}{ k_B T_v}\right)  + \exp\left[-\cfrac{\epsilon_d^{\mx}}{k_B T_{rot}} +\epsilon_d^{\mx} \cfrac{\hat{\delta}_{rot}}{k_B \theta_{rot}} \right] \right. \\ \left. \times \hat{\gn}' \left(-\frac{1}{ k_B T_v}+ \cfrac{1}{k_B T_{rot}} -\cfrac{\hat{\delta}_{rot}}{k_B \theta_{rot}}\right)\right \} 
    \label{avgevhat}
    \end{split}
\end{equation}

\end{widetext}

\bibliography{Bib_CFD_DSMC}
\end{document}